\documentclass[12pt]{article}
\usepackage{epsfig}
\usepackage{amsmath}
\usepackage{hhline}
\usepackage{amssymb}
\usepackage{times}
\usepackage{cite}
\usepackage{graphicx}
\usepackage{ifpdf}
\ifpdf
   \usepackage{epstopdf}
   \usepackage{pdfsync}
\fi
\newlength{\dinwidth}
\newlength{\dinmargin}
\begin{document}
% The rest
\newcommand{\pom}{{I\!\!P}}
\newcommand{\reg}{{I\!\!R}}
\def\gsim{\,\lower.25ex\hbox{$\scriptstyle\sim$}\kern-1.30ex%
\raise 0.55ex\hbox{$\scriptstyle >$}\,}
\def\lsim{\,\lower.25ex\hbox{$\scriptstyle\sim$}\kern-1.30ex%
\raise 0.55ex\hbox{$\scriptstyle <$}\,}
\newcommand{\trm}{m_{\perp}}
\newcommand{\trp}{p_{\perp}}
\newcommand{\trmm}{m_{\perp}^2}
\newcommand{\trpp}{p_{\perp}^2}
\newcommand{\alp}{\alpha_s}
\newcommand{\alps}{\alpha_s}
\newcommand{\sqrts}{$\sqrt{s}$}
\newcommand{\PT}{p_{\perp}}
\newcommand{\JPSI}{J/\psi}
\newcommand{\PO}{I\!\!P}
\newcommand{\xbj}{x}
\newcommand{\xpom}{x_{\PO}}
\newcommand{\dgr}{^\circ}
\newcommand{\gev}{\,\mbox{GeV}}
\newcommand{\GeV}{\mathrm{GeV}}
\newcommand{\mrad}{\mathrm{mrad}}
\newcommand{\GeVS}{\rm GeV^2}
\newcommand{\xp}{x_p}
\newcommand{\xpi}{x_\pi}
\newcommand{\xg}{x_\gamma}
\newcommand{\xgj}{x_\gamma^{jet}}
\newcommand{\xpj}{x_p^{jet}}
\newcommand{\xpij}{x_\pi^{jet}}
\renewcommand{\deg}{^\circ}
\newcommand{\qsq}{\ensuremath{Q^2} }
\newcommand{\et}{\ensuremath{E_t^*} }
\newcommand{\rap}{\ensuremath{\eta^*} }
\newcommand{\gp}{\ensuremath{\gamma^*}p }
\newcommand{\der}{{\mathrm d}}
% Journal macro
\def\Journal#1#2#3#4{{#1} {\bf #2}, #4 (#3)}
\def\NCA{\em Nuovo Cimento}
\def\NIM{\em Nucl. Instrum. Methods}
\def\NIMA{{\em Nucl. Instrum. Methods} {\bf A}}
\def\NPB{{\em Nucl. Phys.}   {\bf B}}
\def\PLB{{\em Phys. Lett.}   {\bf B}}
\def\PRL{\em Phys. Rev. Lett.}
\def\PRD{{\em Phys. Rev.}    {\bf D}}
\def\PR{{\em Phys. Rev.}    }
\def\ZPC{{\em Z. Phys.}      {\bf C}}
\def\ZP{{\em Z. Phys.}      }
\def\EJC{{\em Eur. Phys. J.} {\bf C}}
\def\EJA{{\em Eur. Phys. J.} {\bf A}}
\def\CPC{\em Comp. Phys. Commun.}
\begin{titlepage}

\begin{flushleft}
{\tt DESY 11-093    \hfill    ISSN 0418-9833} \\
{\tt June 2011}                              \\
\end{flushleft}

\vspace*{30mm}

\rm

\begin{center}
\begin{Large}
{\boldmath \bf
Measurement of Photon Production in the Very Forward Direction
in Deep-Inelastic Scattering at HERA
}

\vspace{2cm}
\rm
H1 Collaboration

\end{Large}
\end{center}

\vspace{2cm}

\begin{abstract}
\noindent
\rm The production of photons at very
small angles with respect to the proton beam direction
is studied in deep-inelastic positron-proton scattering  at HERA.
The data are taken with the H1 detector in the years 2006 and 2007 and
correspond to an integrated luminosity of $126~\mathrm{pb}^{-1}$.
The analysis covers the range of
negative four momentum transfer squared at the positron vertex
$6<Q^2<100$~GeV$^2$ and inelasticity $0.05<y<0.6$.
Cross sections are measured for the most energetic
photon with pseudorapidity $\eta>7.9$ as a function of its 
transverse momentum $p_T^{lead}$ and longitudinal momentum fraction
of the incoming proton $x_L^{lead}$.
In addition, the cross sections are studied as a function of
the sum of the longitudinal momentum fraction $x_L^{sum}$
of all photons in the pseudorapidity range $\eta>7.9$.
The cross sections are normalised to the inclusive deep-inelastic
scattering cross section and compared to the predictions
of models of deep-inelastic scattering and models of the hadronic interactions
of high energy cosmic rays.
\end{abstract}

\vspace*{10mm}

\it \normalsize
\begin{center} Submitted to \EJC
\end{center}

\end{titlepage}
%----------------------------------------------------------------------------------
%-- H1AUTS Author list by names 
%-- Status: Wed Jun  1 14:13:45 CEST 2011  Number of authors = 203 

\noindent
F.D.~Aaron$^{5,48}$,           %BUCH-PD        11/06           Aaron               
C.~Alexa$^{5}$,                %BUCH-PD        06/06           Alexa               
V.~Andreev$^{25}$,             %LPI -PD        8/88            Andreev             
S.~Backovic$^{30}$,            %PODG-PD        03/02           Backovic            
A.~Baghdasaryan$^{38}$,        %YERE-PD        09/03           Baghdasaryana       
S.~Baghdasaryan$^{38}$,        %YERE-ST        02/10           Baghdasaryans       
E.~Barrelet$^{29}$,            %PARI-PD        11/99           Barrelet            
W.~Bartel$^{11}$,              %DESY-PD        8/88            Bartel              
K.~Begzsuren$^{35}$,           %ULBA-PD        04/06           Begzsuren           
A.~Belousov$^{25}$
% ...

%-- H1 Institutes 
\bigskip\noindent{\it
% ...
$ ^{2}$ Vinca Institute of Nuclear Sciences, University of Belgrade,
          1100 Belgrade, Serbia \\
% ...
$ ^{5}$ National Institute for Physics and Nuclear Engineering (NIPNE) ,
          Bucharest, Romania$^{ m}$ \\
% ...
$ ^{11}$ DESY, Hamburg, Germany \\
% ...
$ ^{25}$ Lebedev Physical Institute, Moscow, Russia$^{ e}$ \\
% ...
$ ^{29}$ LPNHE, Universit\'e Pierre et Marie Curie Paris 6,
          Universit\'e Denis Diderot Paris 7, CNRS/IN2P3, Paris, France \\
 $ ^{30}$ Faculty of Science, University of Montenegro,
          Podgorica, Montenegro$^{ n}$ \\
% ...
$ ^{35}$ Institute of Physics and Technology of the Mongolian
          Academy of Sciences, Ulaanbaatar, Mongolia \\
% ...
$ ^{38}$ Yerevan Physics Institute, Yerevan, Armenia \\
% ...
$ ^{48}$ Also at Faculty of Physics, University of Bucharest,
          Bucharest, Romania \\
% ...
$ ^e$ Supported by the Deutsche Forschungsgemeinschaft \\
% ...
$ ^m$ Supported by the Romanian National Authority for Scientific Research
      under the contract PN 09370101 \\
 $ ^n$ Partially Supported by Ministry of Science of Montenegro,
      no. 05-1/3-3352 \\
}
	
\newpage
%----------------------------------------------------------------------------------

\section{Introduction}

\noindent
%--------------------------------------------------------------------------------------
Measurements of particle production at very small
angles with respect to the proton beam direction (forward direction)
in positron-proton collisions are important for the %theoretical
understanding of proton fragmentation.
These measurements %of forward particle production 
also provide important constraints for the modelling of the
high energy air showers and thereby are very valuable for the
understanding of high energy cosmic ray data.
The H1 and ZEUS experiments at the
$e^{\pm}p$ collider HERA have
published several analyses on the production of forward
protons and neutrons which carry  a large fraction of the longitudinal
momentum of the incoming proton
\cite{Adloff:1998yg,Chekanov:2002pf,Chekanov:2007tv,Chekanov:2008tn,Aaron:2010ze}.
These measurements probe different mechanisms related to the baryon production
in forward direction, such as elastic scattering of the proton, 
diffractive dissociation, pion exchange and string fragmentation.
In particular, these measurements  test the
hypothesis of limiting fragmentation~\cite{Benecke:1969sh,Chou:1994dh},
according to which, in the high-energy limit, the cross
section for the inclusive production of particles in the target
fragmentation region is independent of the incident projectile energy.  
This hypothesis implies, that in deep-inelastic scattering (DIS) forward particle 
production cross sections are independent of the Bjorken-$x$ and the 
virtuality of the exchanged photon $Q^2$.

The measurement of the photon production in the 
forward direction can provide new input
to the understanding of proton fragmentation, and is complementary
to forward baryon measurements.
The production of photons and $\pi^0$ mesons in the proton fragmentation 
region has been studied in $\bar{p}p$ and $pp$ collisions
 at SPS and the LHC colliders~~\cite{Pare:1989mr,Adriani:2011nf}.
The analysis presented here is the first measurement of very forward
photons in DIS $e^+p$ collisions at HERA.  
The photons are detected at very small angles below $0.75$~mrad 
with respect to proton beam direction.
It relies on the upgraded H1 Forward Neutron Calorimeter (FNC)  which includes an
electromagnetic section.

\section{Experimental Procedure and Data Analysis}
%--------------------------------------------------------------------------------------
The data used in this analysis were collected with the H1 detector at
HERA in the years 2006 and 2007 and correspond to an integrated
luminosity of $126~ \rm pb^{-1}$. %For the period used in this analysis,
During the period corresponding to the analysis data set
HERA collided positrons and protons with energies of $E_e=27.6$~GeV
and $E_p=920$~GeV, respectively, corresponding to a
centre-of-mass energy of $\sqrt{s}=319$~GeV.

\subsection{H1 detector}

A detailed description of the H1 detector can be found elsewhere
\cite{Abt:1996hi,Abt:1996xv,Appuhn:1996na,Pitzl:2000wz,spacaltest,Andrieu:1993kh}.
Only the detector components relevant to this analysis are briefly
described here.
The origin of the right-handed
 H1 coordinate system is the nominal $e^+p$ interaction point. 
 The direction of the proton beam defines  the positive $z$ axis; 
 the polar angle $\theta$ is measured with respect to  this axis. 
 Transverse momenta are measured in the $x$--$y$ plane.
 The pseudorapidity is defined by $\eta =  -\ln{(\tan\frac{\theta}{2})}$ 
 and is measured in the lab frame.
 The polar angles $\theta<0.75$~mrad correspond to pseudorapidity 
 range $\eta>7.9$.

The interaction region is surrounded by a two-layer silicon
strip detector and two large concentric drift chambers.
% , operated inside a $1.16$~T solenoidal magnetic field.
Charged particle momenta are measured in the
angular range \break \mbox{$25\deg<\theta<155\deg$}.
 The tracking system is surrounded by a finely segmented
Liquid Argon (LAr) calorimeter, which covers the polar angle range of
\mbox{$4\deg<\theta<154\deg$} with full azimuthal acceptance.  The LAr
calorimeter consists of an electromagnetic section with lead absorber
and a hadronic section with steel absorber.  The total depth of the
LAr calorimeter ranges from $4.5$ to $8$ hadronic interaction lengths.
The backward region ($153\deg<\theta<177.8\deg$) is covered by a
lead/scintillating-fibre calorimeter (SpaCal).
Its main purpose is the detection of the scattered positron. 
% The polar angle of the positron is measured with a precision of $1$~mrad.
The energy resolution for positrons is 
$\sigma(E)/E\approx 7.1\%/\sqrt{E[\GeV]}\oplus 1\%$,
as determined in test beam measurements~\cite{spacaltest}.
The LAr and SpaCal
calorimeters are surrounded by a superconducting solenoid which
provides a uniform magnetic field of $1.16$~T along the beam direction.

The luminosity is measured via the Bethe-Heitler Bremsstrahlung
process $ep \rightarrow e'p \gamma$, the final state photon being detected in a
tungsten/quartz-fibre sampling calorimeter at $z=-103$~m.

The data sample of this analysis was collected using
 triggers which require the scattered positron to be measured in the SpaCal.
 The trigger efficiency is about $96\%$ for the analysis phase space
 as determined from data using independently triggered data.

 \begin{figure}[h]
 \hspace*{40mm}{\large \bf (a)} \hspace*{60mm} {\large \bf (b)}
 \vspace*{-2mm}

   \epsfig{file=desy-11-093.fig1a.eps,width=68mm}
   \hspace*{20mm}
   \epsfig{file=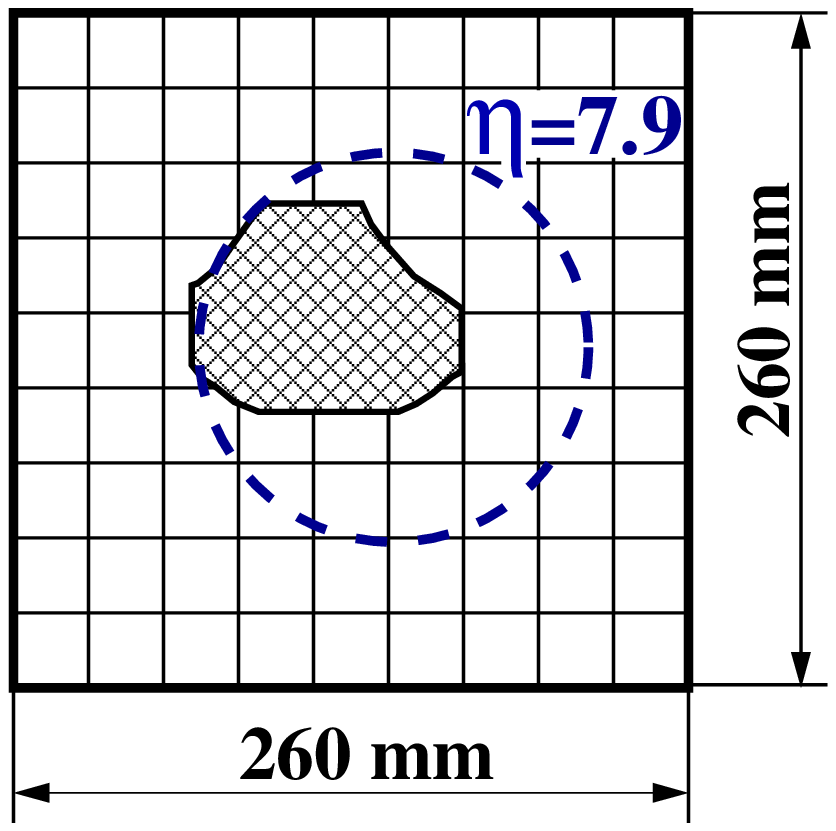,width=58mm}
   \caption{(a) A schematic view of the H1 FNC.
     (b) Layout of $9$ vertical and $9$ horizontal readout strips of 
     the Preshower Calorimeter.
     The hatched area shows the geometrical acceptance window defined
     by the beam-line elements.
    The area corresponding to $\eta>7.9$ is indicated by dashed circle.
    }
   \label{gen_view}
 \end{figure}

%--------------------------------------------------------------------------------------
\subsection{Detection of forward neutral particles}
\label{sec:fncdet}

Neutral particles produced at very small polar angles 
can be detected in the FNC calorimeter, which is situated 
at a polar angle of $0\deg$, at $z=+106$~m from the interaction point.
A schematic view of the H1 FNC used during the HERA-II running period
is shown in figure~\ref{gen_view}a.
A detailed description
of the detector is given in \cite{Aaron:2010ze}.  The FNC consists of the Main
Calorimeter and the Preshower Calorimeter.
The Main Calorimeter is a lead-scintillator 
sandwich calorimeter with a total length of $8.9$ nuclear interaction lengths.
The Preshower Calorimeter is a $40$~cm long lead-scintillator
sandwich calorimeter.  The length corresponds to about $60$ radiation
lengths.  The Preshower Calorimeter is composed of $24$ planes: the
first $12$ planes each consist of a lead plate of $7.5$~mm thickness
and a scintillator plate of $2.6$~mm thickness.  The second $12$ planes
each consist of a lead plate of $14$~mm thickness and a scintillator
plate of $5.2$~mm thickness.
The transverse size of the scintillating plates is \mbox{$26\times26~\rm cm^2$}.
Each scintillating plate has $45$ grooves
with $1.2$~mm wavelength shifter  fibres attached down one side.
The orientation of fibres alternates from horizontal to vertical in
consecutive planes. For each plane, the fibres are bundled into nine
strips of five fibres each. Longitudinally,  all strips are combined leading to 
$9$ vertical and $9$ horizontal towers which are finally connected to
$18$ photomultipliers.

The acceptance of the FNC is defined by the aperture of the HERA
beam-line magnets and is limited to scattering angles of
$\theta\lsim 0.8$~mrad with approximately $30\%$ azimuthal coverage.  The
geometrical acceptance window of the FNC is shown in
figure~\ref{gen_view}b together with the layout of the Preshower Calorimeter
readout strips.

The longitudinal segmentation of the FNC allows efficient
discrimination of photons from hadrons.  The photon reconstruction
algorithm is based on the fact that electromagnetic showers are
fully contained in the Preshower Calorimeter with no energy deposits
 above the noise level in the Main Calorimeter.
 For high energy neutrons most of the energy is contained in the
Main Calorimeter.  However, low energy neutrons deposit large
fractions of their energy in the Preshower Calorimeter.  
The fraction of neutrons which can be misidentified as photons is 
about $10\%$ for $90$~GeV neutrons decreasing to
below $1\%$ for neutrons with an energy of $200$~GeV, as determined
from the Monte Carlo (MC) simulation.
The energy deposits in the FNC which are contained in the
Preshower Calorimeter are classified as electromagnetic clusters
and are considered as photon candidates.
The detection and reconstruction efficiency for photons in the
measured angular range $\theta<0.75$~mrad, as estimated from MC 
simulation, is about $85\%$ for $100$~GeV photons
increasing to $95\%$ for photons with energies of $900$~GeV.
Losses are mainly due to interactions with the beampipe.

%------------
% Calibration
%-------------
%
All modules of the FNC were initially calibrated at CERN using
$120$-$230$~GeV electron and $120$-$350$~GeV hadron beams.  After the
calorimeter was installed at DESY, the stability of
calibration constants was monitored using interactions between the
proton beam and residual gas in the beam pipe, as described in
\cite{Aaron:2010ze}. Refined calibration constants for
electromagnetic showers are determined using an iterative procedure
based on the assumption that the maximum photon energy, $E_\gamma^{max}$,
as measured in the Preshower Calorimeter,
is expected to be equal to the proton beam energy in case of unlimited statistics.
This calibration procedure also utilises data from HERA runs with
reduced proton beam energies of $460$~GeV and $575$~GeV.
The validity of this algorithm is tested with MC simulation.

\begin{figure}[t]
\epsfig{file=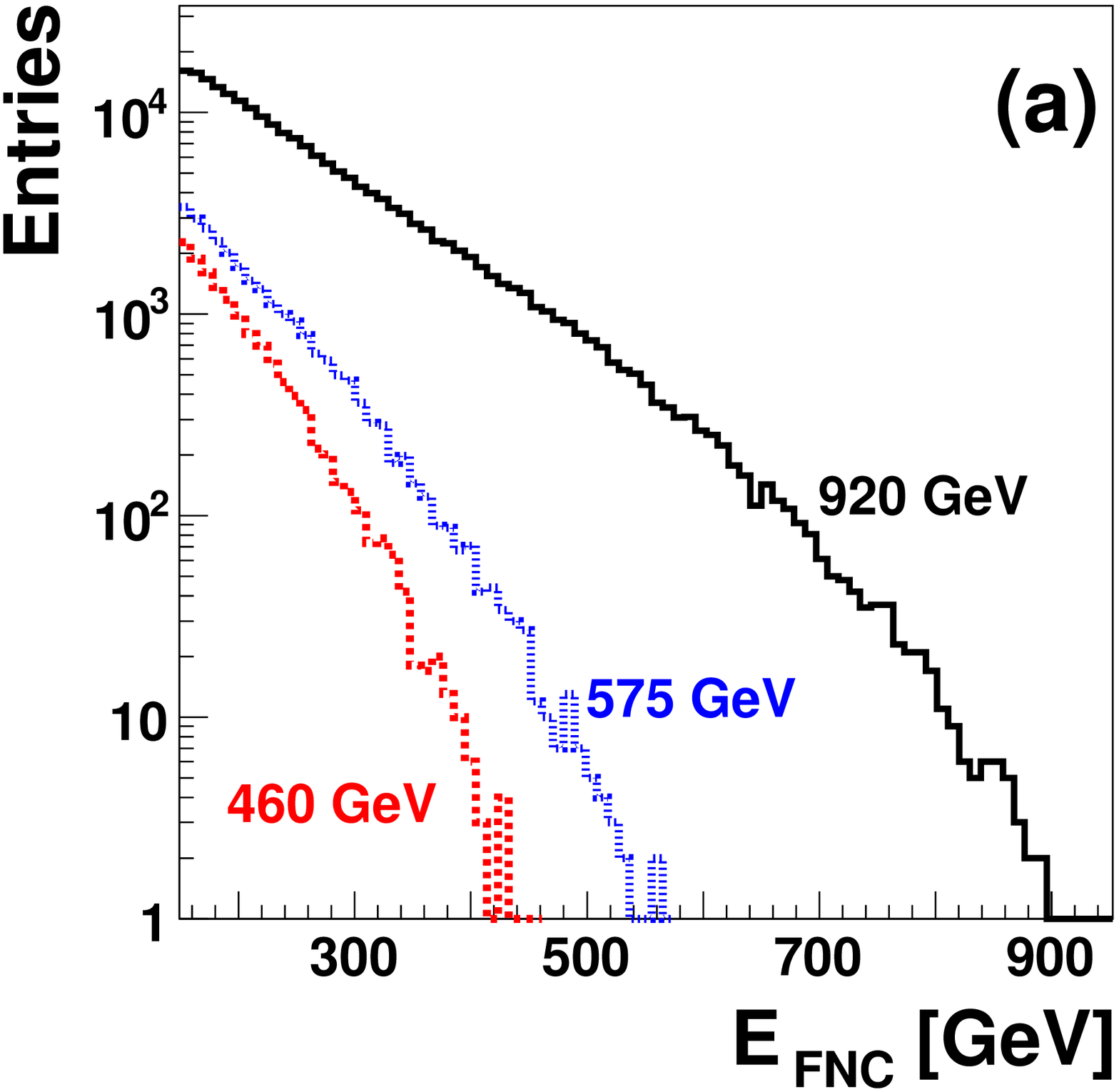,width=77mm}
%\hspace*{-.5cm}\epsfig{file=desy-11-093.fig2b.eps,width=77mm}
%\epsfig{file=desy-11-093.fig2c.eps,width=77mm}
\epsfig{file=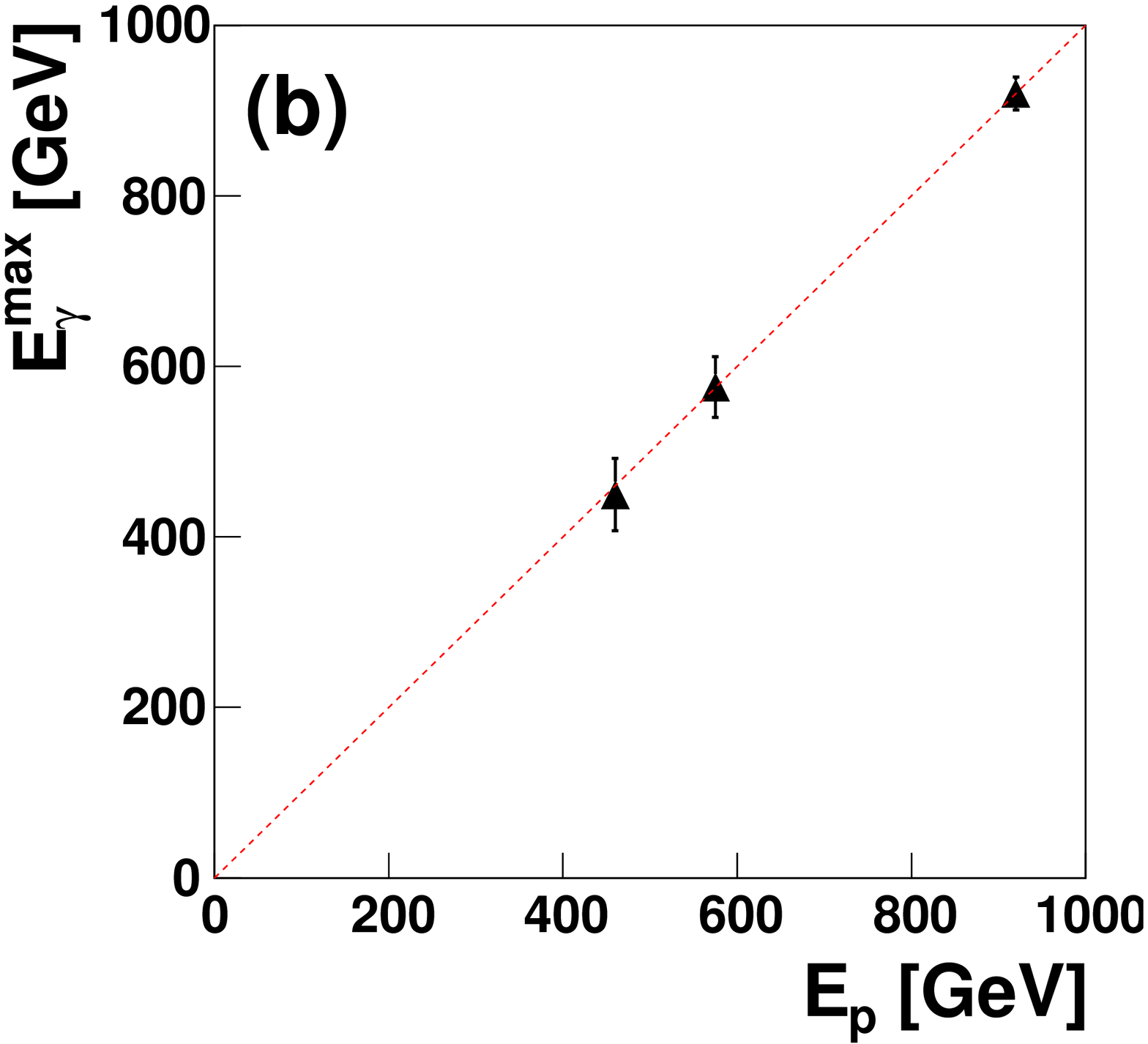,width=78mm}
\caption{
\label{fig:calibration}
(a)
The measured photon energy spectra for three proton beam energies.
% ($460$, $575$ and $920$~GeV).
%(b)
%The measured photon energy spectra for $E_p=920$~GeV.
%The different lines represent the development 
%of the photon energy distribution when gradually increasing the statistics.
%(c)
% The fitted upper edges of the energy distributions as a function of
% the number of iterations corresponding to the gradual increase of the event statistics.
% Results of the fit with a function $f(i)=E_{\gamma}^{max}+a\cdot exp(b\cdot i)$
% are shown by the dashed lines. 
(b) The correlation between the proton beam energy and 
%the upper edge of photon energy distribution 
$E_{\gamma}^{max}$.
}
\end{figure}

The measured photon energy spectra for the three proton beam energies
are displayed in figure~\ref{fig:calibration}a.
The correlation between the beam
energy and the maximum photon energy $E_{\gamma}^{max}$ as determined by the iterative 
procedure and after applying the calibration
is shown in  figure~\ref{fig:calibration}b.
Using this calibration procedure, the linearity of the energy response
and the absolute energy scale are verified to a precision of $5\%$.  

%
%Figures~\ref{fig:calibration}b shows the 
%photon energy spectra for $E_{p}=920$~GeV,
%where the different lines  represent the development of 
%the distribution when gradually increasing the event statistics.
%It can be seen, that the upper edge of the distribution
%approaches the beam energy with the increase of statistics.
%Fitting these distributions, the values of the upper edges,
%$E_{\gamma}^{edge}$, are determined.
%In figure~\ref{fig:calibration}c the values of $E_{\gamma}^{edge}$
%are shown as a function of the number of iterations
%corresponding to the gradual raise of statistics
%for three proton beam energies.
%Fitting the dependence on the number of iterations with a function
%$f(i)=E_\gamma^{max}+a\cdot exp(b\cdot i)$, where $i$ is the number
%of iterations, the upper limits of
%the photon energy distributions, $E_{\gamma}^{max}$,
%are determined and the calibration constants are derived.
%The validity of this algorithm is tested with MC simulation.
%

The energy resolution of the FNC calorimeter for electromagnetic
showers is $\sigma(E)/E \approx 20\%/\sqrt{E~[\GeV]} \oplus 2\%$,
as determined in test beam measurements.
The spatial resolution for single electromagnetic showers and for those
hadronic showers which started to develop in the Preshower Calorimeter
is about $2$~mm.
%--------------------------------------------------------------------------------------
\subsection{Kinematics and event selection }

The kinematic variables used to describe high energy DIS interactions are
the exchanged photon virtuality $Q^2$, the inelasticity $y$ and 
the Bjorken scaling variable $x_{Bj}$. They are defined as
\begin{equation}
Q^2=-q^2\; , \hspace{3em}  
x_{Bj}=\frac{Q^2}{2 p \cdot q}\; , \hspace{3em}  
y=\frac{p \cdot q}{p \cdot k}\;,
\end{equation}
where $p$, $k$ and $q$ are the four-momenta of the incident proton,
the incident positron and the virtual photon, respectively.

The selection of DIS events is based on the identification
of the scattered positron as the most energetic compact calorimetric deposit
in the SpaCal  with an
energy $E_e'>11$~GeV and a polar angle $156\deg<\theta_e'<175\deg$.
The \mbox{$z$ coordinate} of the primary event vertex is required to be
within $\pm 35$~cm of the nominal position of the interaction point.
The hadronic final state is reconstructed using an energy flow algorithm 
which combines charged particles measured in the tracker with information 
from the SpaCal and LAr calorimeters~\cite{Peez:2003zd,Hellwig:2004yp}.
To suppress events with hard initial state radiation, as well as
events originating  from  non-$e^+p$ interactions,
the quantity $\sum{E - p_z}$, summed over
all reconstructed final state particles including the positron, 
is required to lie between  $35$~GeV and $70$~GeV.
This quantity, which uses the energy and longitudinal momentum
component of each final state particle, is expected to be twice the
electron beam energy for fully contained events.
In addition, events are restricted to
the kinematic range $6<Q^2<100$~GeV$^2$ and $0.05<y<0.6$.
These variables are reconstructed using a technique
which optimises the resolution throughout the
measured $y$ range, exploiting information from both the scattered
positron and the hadronic final state \cite{Adloff:1997sc}.
The data sample of the DIS events contains about $9.2$~million events.

Events containing forward photons are selected by requiring an
electromagnetic cluster in the FNC with a pseudorapidity above $7.9$
and an energy above $92$~GeV, which corresponds to the longitudinal 
momentum fraction $x_L=E_\gamma/E_p>0.1$,
 where $E_p$ and $E_\gamma$ are
the proton beam and forward photon energy, respectively.
The data sample contains $78740$ events.

In this analysis %the
normalised
differential cross sections are measured for the most energetic
forward photon (leading photon) with pseudorapidity
$\eta>7.9$ as a function of
its longitudinal momentum fraction $x_L^{lead}$ and transverse momentum
$p_T^{lead}$, in the range  $0.1<x_L^{lead}<0.7$.
Cross sections are also measured as a function of the sum  of 
longitudinal momentum fractions of 
all forward photons with $\eta>7.9$,
$x_L^{sum}=\sum x_L$, in the range $0.1<x_L^{sum}<0.95$.
%The determination of 
These cross sections %is performed 
are given as the
fraction of DIS events 
%which have the 
having forward photon in the $\eta$--$x_L$ regions given above.
Finally, the ratio of the forward photon production
cross section to the inclusive DIS cross section is 
presented as a function of $Q^2$ and $x_{Bj}$.

%---------------------------------------------------------------------------
\subsection{Monte Carlo simulations and corrections to the data}
\label{sec:mc}

Monte Carlo simulations are used to correct the data for the
effects of detector acceptance, inefficiencies, migrations between
measurement bins due to finite detector resolution and QED
radiation from the positron.  All generated events are passed through
a GEANT3~\cite{Brun:1978fy} based simulation of the H1 apparatus and
are then processed using the same reconstruction and analysis chain as
is used for the data.

The DJANGOH~\cite{Charchula:1994kf}
program is used to generates inclusive DIS events.  It is based on leading order
electroweak cross sections and takes into account QCD effects up to
order $\alpha_s$.
Higher order QCD effects are simulated using
leading log parton showers as implemented in LEPTO~\cite{Ingelman:1996mq},
or using the Colour Dipole Model (CDM) as implemented in ARIADNE
\cite{Lonnblad:1992tz}.  Subsequent hadronisation effects are modelled
using the Lund string fragmentation model as implemented in JETSET
\cite{Andersson:1983ia,Sjostrand:1995iq}.  
Higher order electroweak processes are
simulated using an interface to HERACLES~\cite{Kwiatkowski:1990es}.
The LEPTO program includes the simulation of soft colour interactions
(SCI)~\cite{Edin:1995gi}, in which the production of
diffraction-like configurations is enhanced via non-perturbative
colour rearrangements between the outgoing partons.
In the measured $x_L$ range, omitting the SCI in the LEPTO would decrease
the predicted yield of forward photons by $5\%$ at lowest $x_L$
and $2\%$ at highest $x_L$.
The simulation with CDM uses the parameters tuned to
describe H1 forward jet measurements \cite{Aktas:2005up}.
The DJANGOH MC simulations are calculated using the H1PDF~2009
parameterisation~\cite{Aaron:2009kv} of the parton distributions in the
proton.  In the following, the predictions based on LEPTO
and ARIADNE are denoted LEPTO and CDM, respectively.

The measurements are also compared with the predictions of several hadronic
interaction models which are commonly used for the simulation of cosmic ray
air shower
cascades: EPOS~\cite{Werner:2005jf}, QGSJET~01~\cite{Kalmykov:1993qe,Kalmykov:1997te},
QGSJET~II~\cite{Ostapchenko:2005nj,Ostapchenko:2007qb}  and
SIBYLL~\cite{Engel:1992vf,Ahn:2009wx}.
These phenomenological models, based on
general principles such as unitarity and analyticity of scattering
amplitudes, are combined with perturbative QCD predictions for
high-$p_{T}$ processes to obtain a description of the final state.
The programs are interfaced with the PHOJET program~\cite{Engel:1995yda}
for the simulation of $e^+p$ interactions.

 In all of these models, the main source of forward photons is the decay
 of $\pi^0$ mesons produced from the hadronisation of the proton remnant.  
The measured distributions may contain background arising from several
sources.  The background from photoproduction processes, where the
positron is scattered into the backward beam-pipe and a particle from
the hadronic final state fakes the positron signature in the SpaCal,
is estimated using the PHOJET MC generator and found to be negligible.
The selected sample may contain background from neutrons reconstructed 
as electromagnetic clusters as explained above.
For cluster energies above $92$~GeV this background is  found to be 
negligible according to the MC simulation.
The background from the random coincidences of DIS events
with a beam-related background signals in the FNC
is estimated by combining DIS events with forward
particles in adjacent bunch-crossings.  It is found to be smaller than
$1\%$. The background contributions are not subtracted from themeasured
cross sections.

%-------------------------------------------------------------------------------

Two or more particles entering the FNC are reconstructed as a single cluster due
to the relatively large size of the FNC readout modules in combination with a
small  geometrical acceptance window.
According to the MC simulation, low energetic clusters reconstructed in the 
FNC mainly originate from single photons.
The contribution from two photons increases almost linearly
from $10\%$ at about $450$~GeV to $80\%$ at $900$~GeV (the contribution 
from three and more photons is below $1\%$).
Therefore, the measurement of the cross section of single photon
production is limited to $x_L<0.7$, while the
measurement of the total forward photon production cross section
is extended to larger $x_L$.

Factors determined from MC are used to correct distributions 
at the level of reconstructed particles
back to the hadron level on a bin-by-bin basis.
These correction factors include the effects of QED radiation from the positron.
For the calculation of the correction factors the simulations
are reweighted to describe the $x_L$ distributions of the data.
The average of the correction factors determined from LEPTO and CDM is used.
The size of the 
correction  factors varies between $2$ and $3.5$ for $x_{L}^{lead}$, between 
$3$ and $4$ for $x_{L}^{sum}$, between $2.5$ and $12$ for $p_{T}^{lead}$
and are about $3.2$ for the $Q^2$ and $x_{Bj}$ distributions.
They are dominated by the non-uniform azimuthal acceptance of the FNC, 
which is about $30\%$ on average.
The bin purities, defined as the fraction
of events reconstructed in a particular bin that originate from
that bin on hadron level,  vary between $75\%$ and $95\%$.

%--------------------------------------------------------------------------------------
\subsection{Systematic uncertainties}

The systematic uncertainties on the cross section measurements are 
determined using MC simulations, by propagating the corresponding
uncertainty sources through the full analysis chain.

As the cross sections are normalised to the inclusive DIS cross
section measured in this ana\-lysis, some important systematic
uncertainties,
such as the trigger efficiency, the luminosity
and the uncertainties related to the reconstruction of the scattered positron
and of the hadronic final state are largely reduced or cancel.
Uncertainties on the measurements of the scattered positron energy
($1\%$) and angle ($1$~mrad), the energy of the hadronic final state
($4\%$), and the uncertainty on the trigger efficiency ($1\%$) lead to
an average combined uncertainty of up to $2\%$.

The absolute electromagnetic energy scale of the FNC is known to a
precision of $5\%$ as described in section~\ref{sec:fncdet}.
 This leads to an uncertainty of $1\%$ on the cross
section measurement at low energies, increasing to $35\%$ for the
largest $x_L$ values. 
The acceptance of the FNC calorimeter is
defined by the interaction point and the geometry of the HERA magnets
and is determined using MC simulations.
The uncertainty of the impact position of the photon on the FNC
is  due to
beam inclination and the uncertainty on the FNC position. 
It is estimated to be $5$~mm.  This results in uncertainties on the 
FNC acceptance determination of up to $15\%$ for the $x_L$ distributions 
and up to $60\%$ for the $p_T^{lead}$ distribution.
These effects are strongly correlated between measurement bins.
For the $Q^2$ and $x_{Bj}$ measurements, these effects lead to
normalisation uncertainty of approximately $7\%$.

The systematic uncertainty arising from the model dependence of the
data correction
is taken as the difference of the corrections calculated
using the LEPTO and CDM models.
The resulting uncertainty on the cross-section increases from $1\%$ to
$6\%$ for the $x_L^{lead}$ and $p_T^{lead}$ distributions, from $2\%$
to $20\%$ for the $x_L^{sum}$ distribution, and from $1\%$ to $2\%$ for
the $Q^2$ and $x_{Bj}$ distributions.
 Using different parton
distribution functions in the MC simulation results in a negligible
change in the cross section.

The systematic errors shown in the figures and table are calculated as
the quadratic sum of all contributions, which may vary from point to
point. The total systematic error for the normalised cross section
measurements ranges between $8\%$ and $18\%$ for $x_L^{lead}$, $6\%$
and $58\%$ for $p_T^{lead}$,  $8\%$ and $44\%$ for $x_L^{sum}$ and
$7\%$ and $8\%$ for $Q^2$ and $x_{Bj}$.

%--------------------------------------------------------------------------------------
\section{Results}

\begin{figure}[p]
\epsfig{file=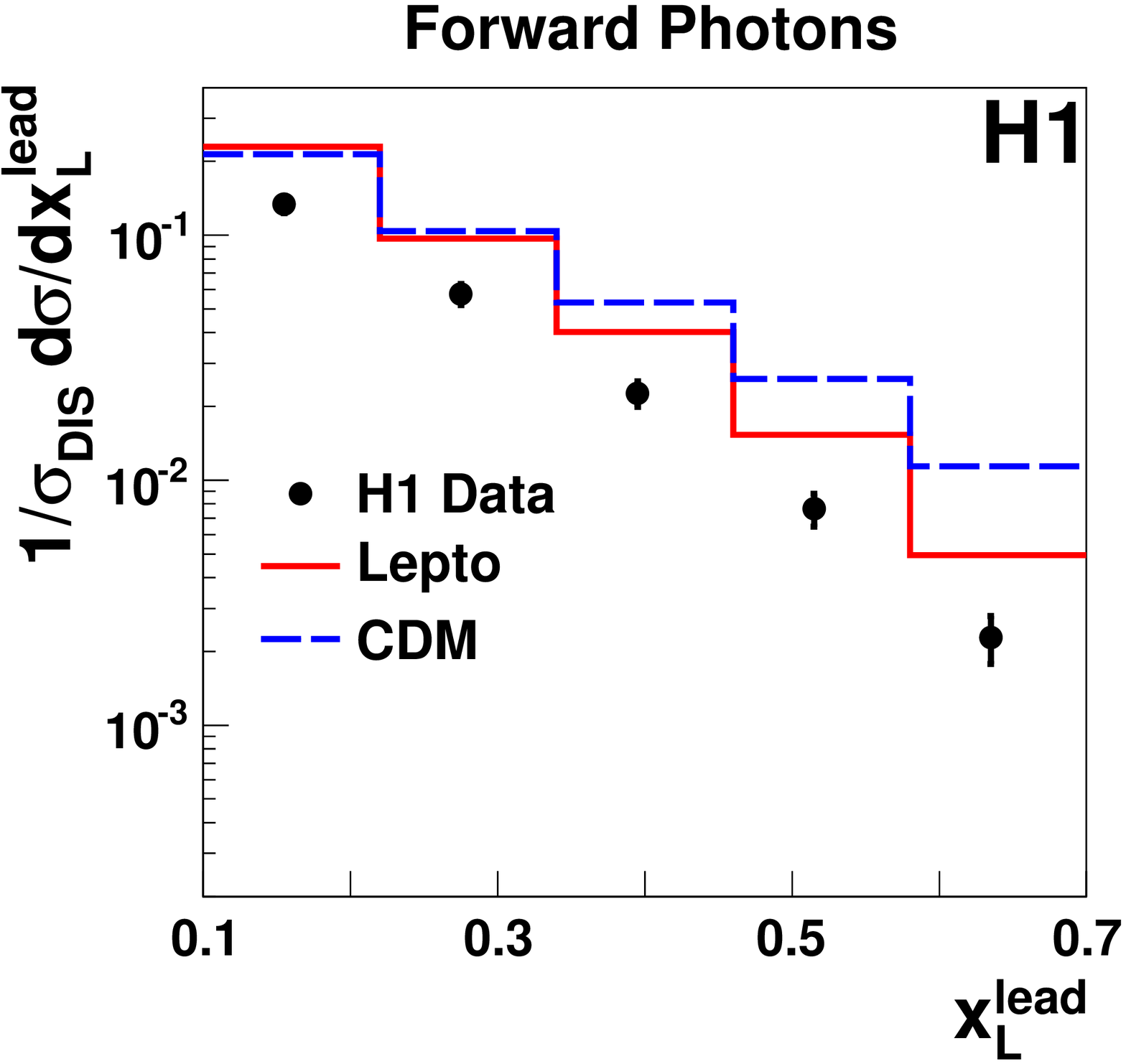,width=80mm}
\epsfig{file=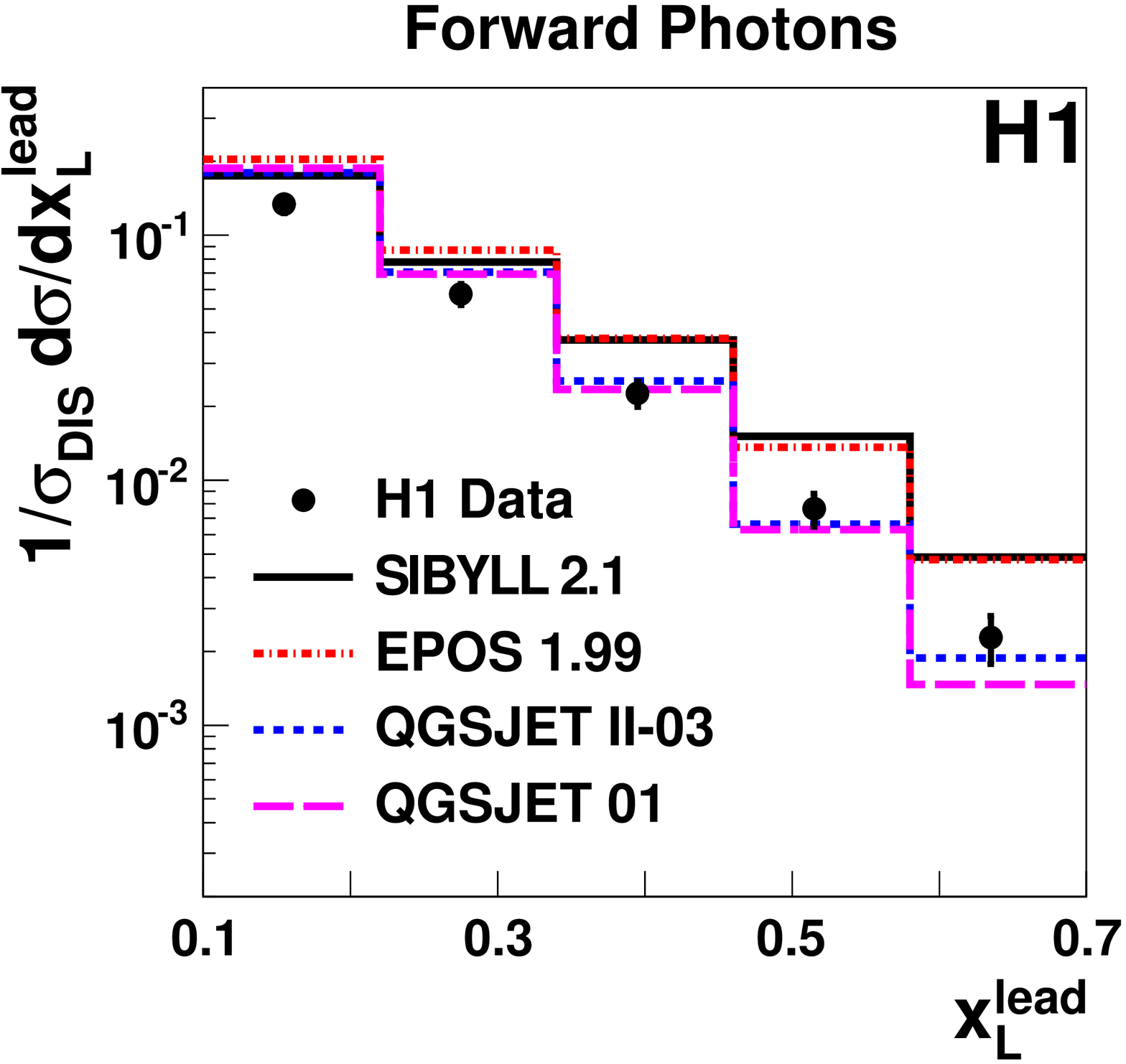,width=80mm}
\epsfig{file=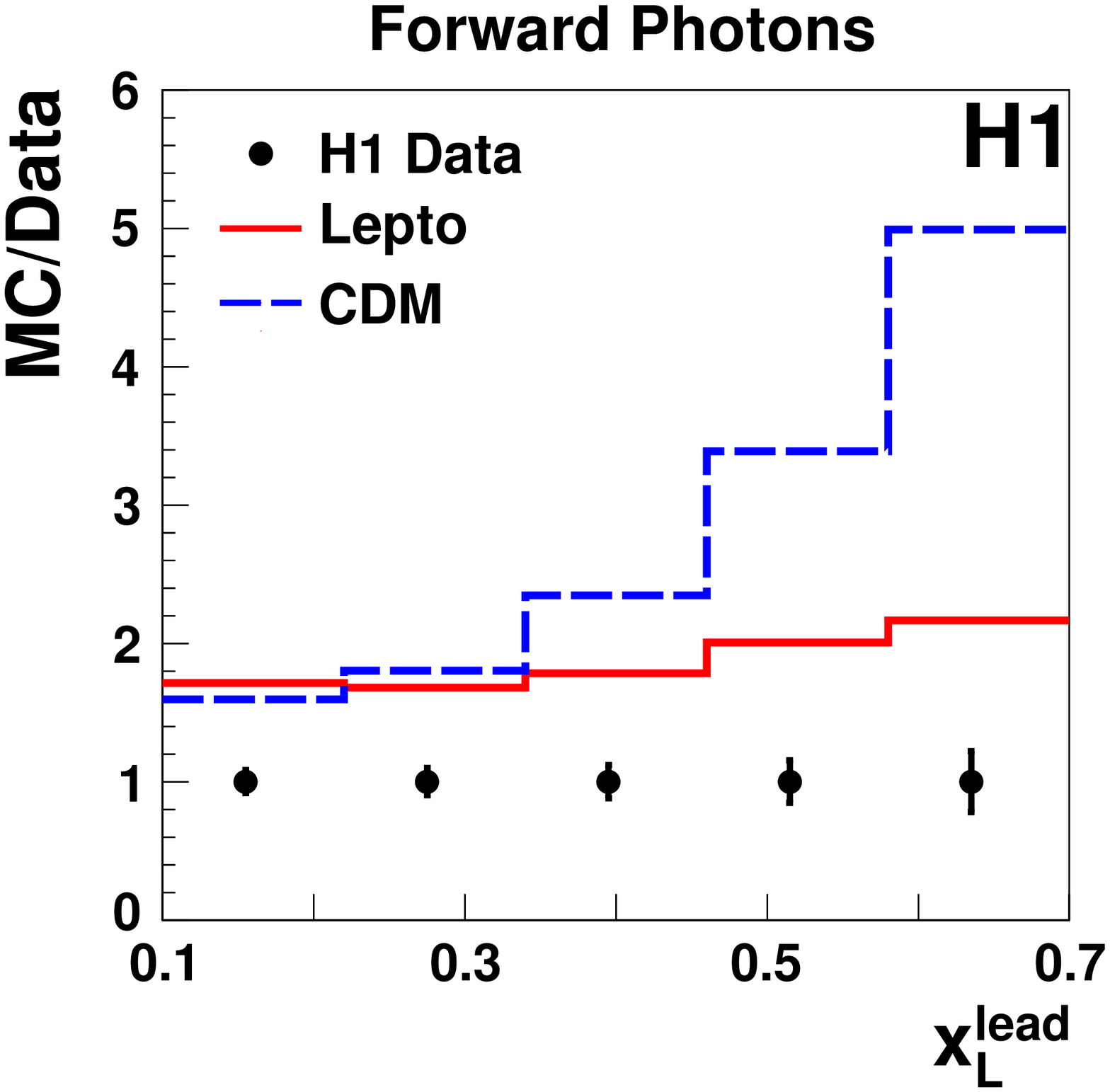,width=80mm}
\epsfig{file=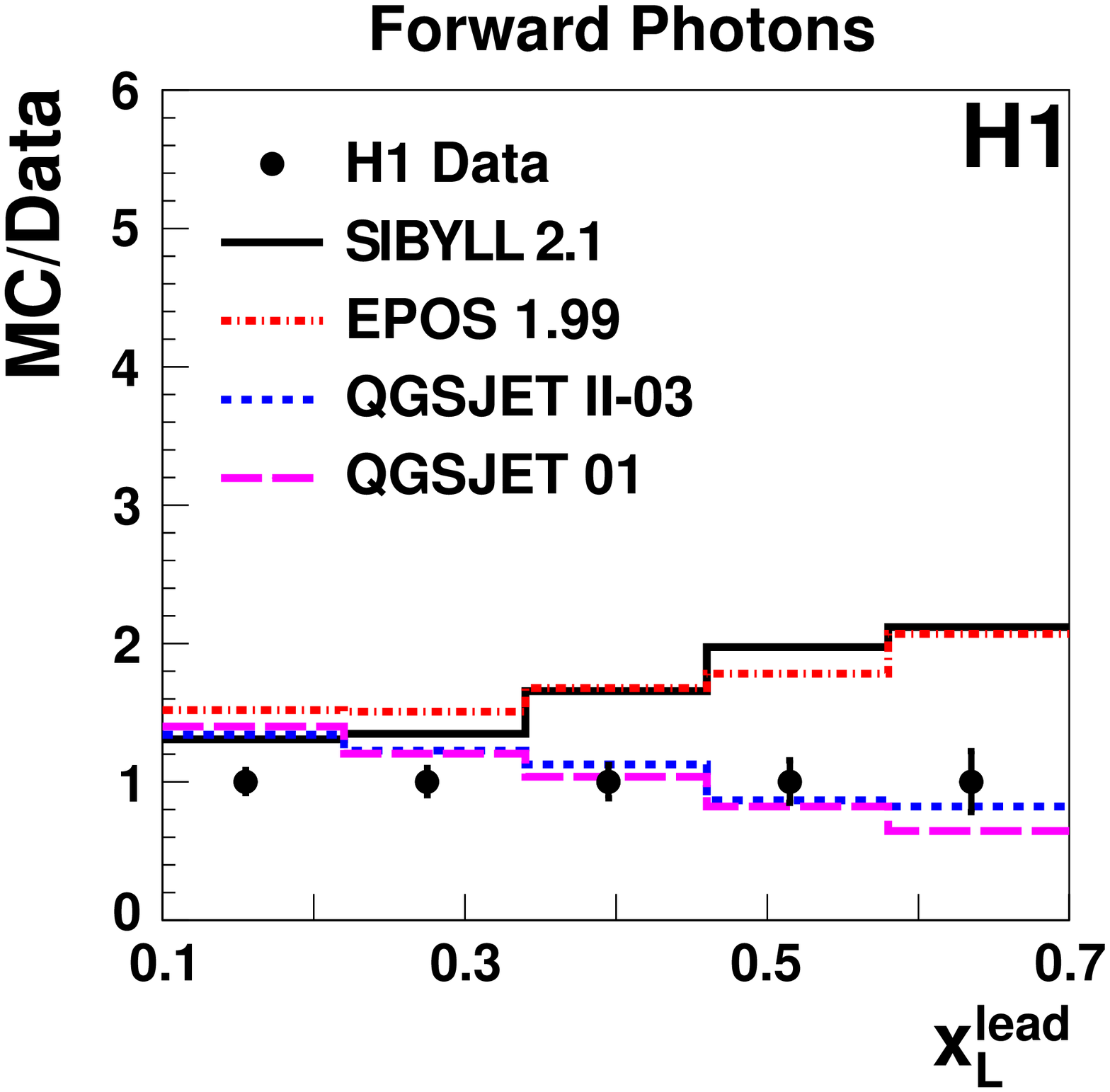,width=80mm}
\caption{
The normalised cross sections for the production of forward photons 
in the pseudorapidity range $\eta>7.9$
in deep-inelastic scattering in the kinematic region $6< Q^2 < 100~$GeV$^2$ and $0.05<y<0.6$
as a function of the longitudinal momentum fraction $x_L^{lead}$ 
of the leading photon
in the range $0.1<x_L^{lead}<0.7$.
The data are compared to two predictions of 
the DJANGOH Monte Carlo simulation, using LEPTO and CDM to simulate higher orders.
Also shown are models of
hadronic interactions, QGSJET, EPOS and SIBYLL.
The lower row shows the ratios of the Monte Carlo predictions to the data. 
The error bars show the total experimental uncertainty, defined as the quadratic sum
of the statistical and systematic uncertainties.
}
\label{fig:CrsecMClead}
\end{figure}

\begin{figure}[p]
\epsfig{file=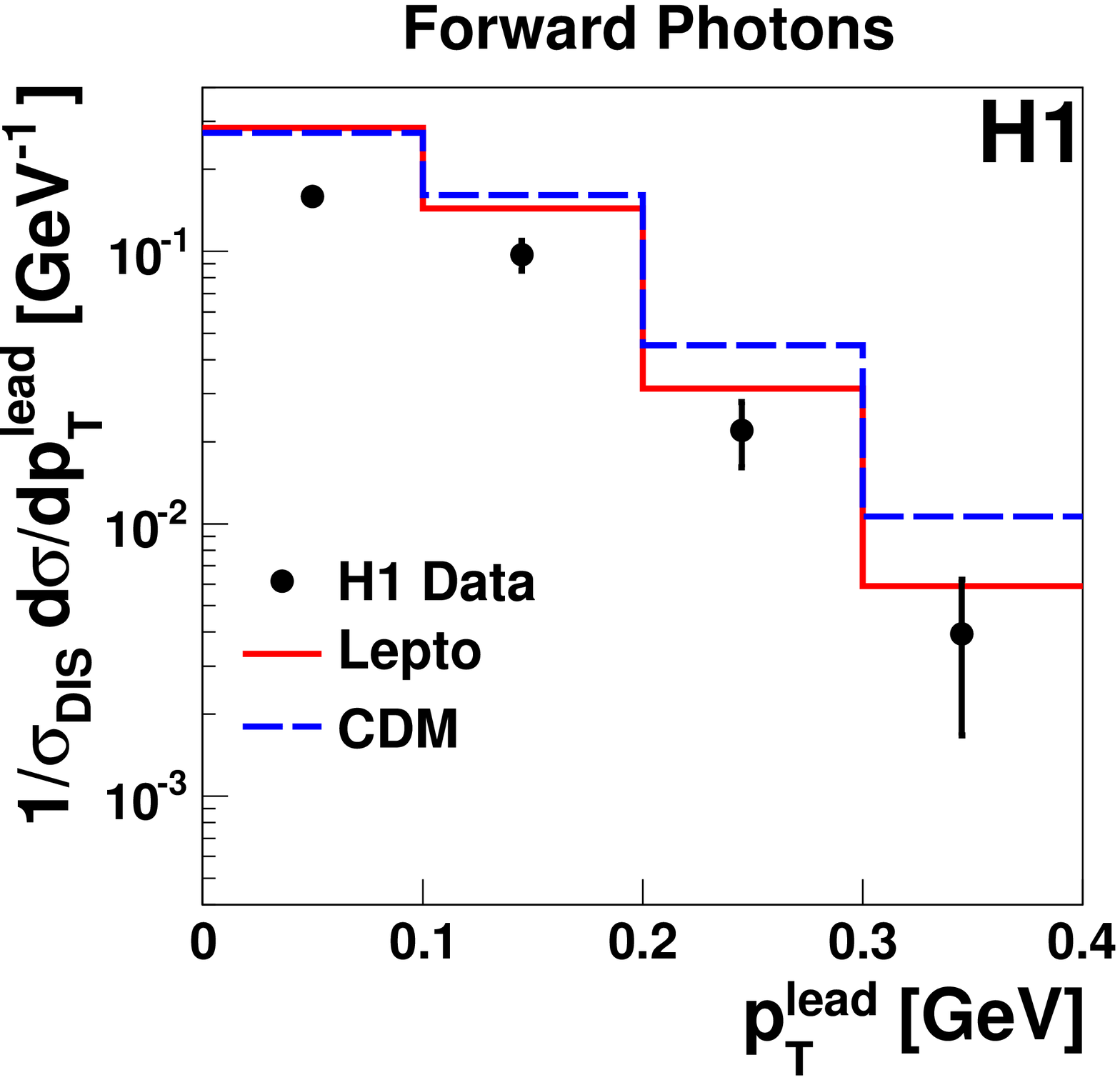,width=80mm}
\epsfig{file=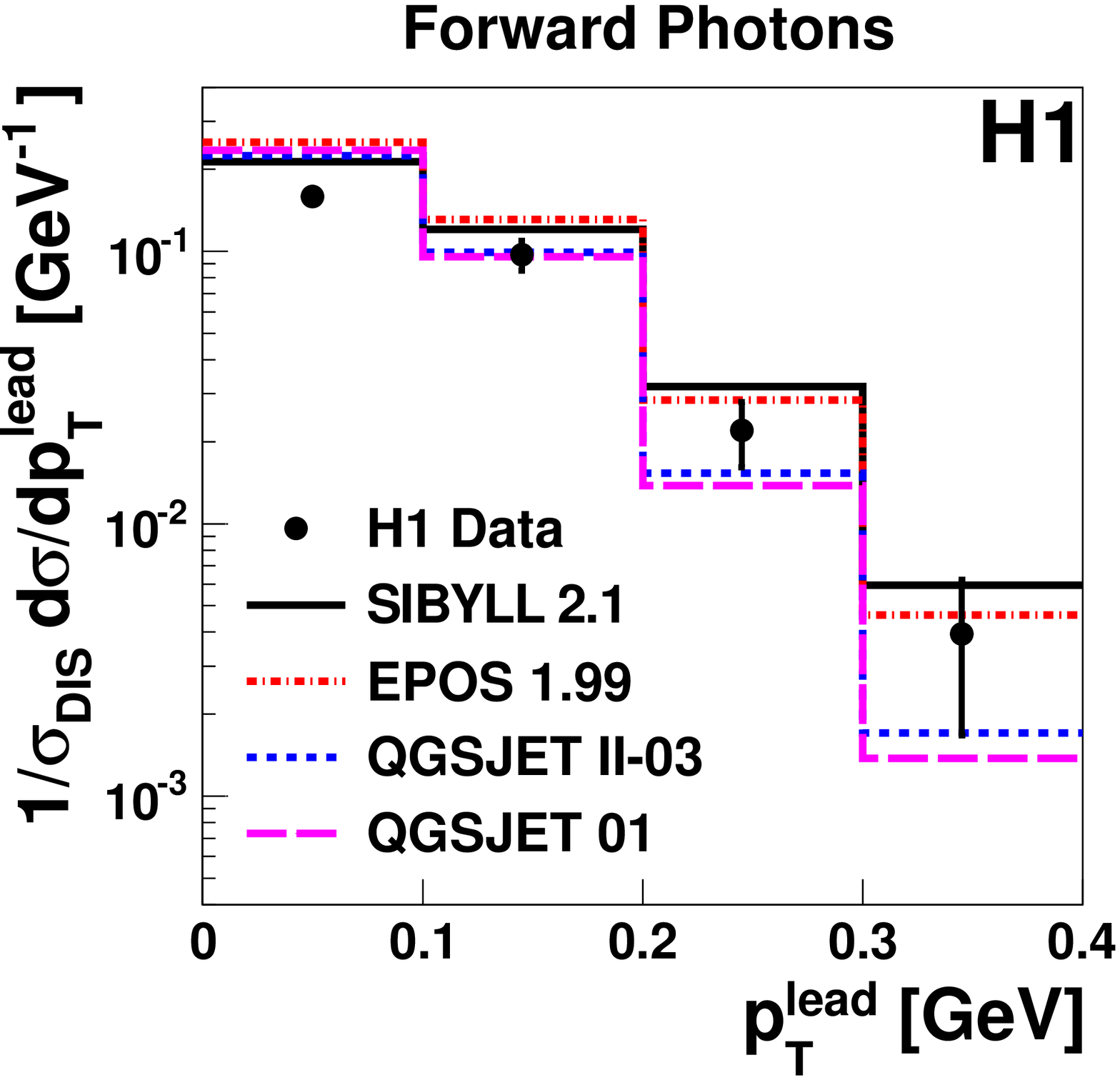,width=80mm}
\epsfig{file=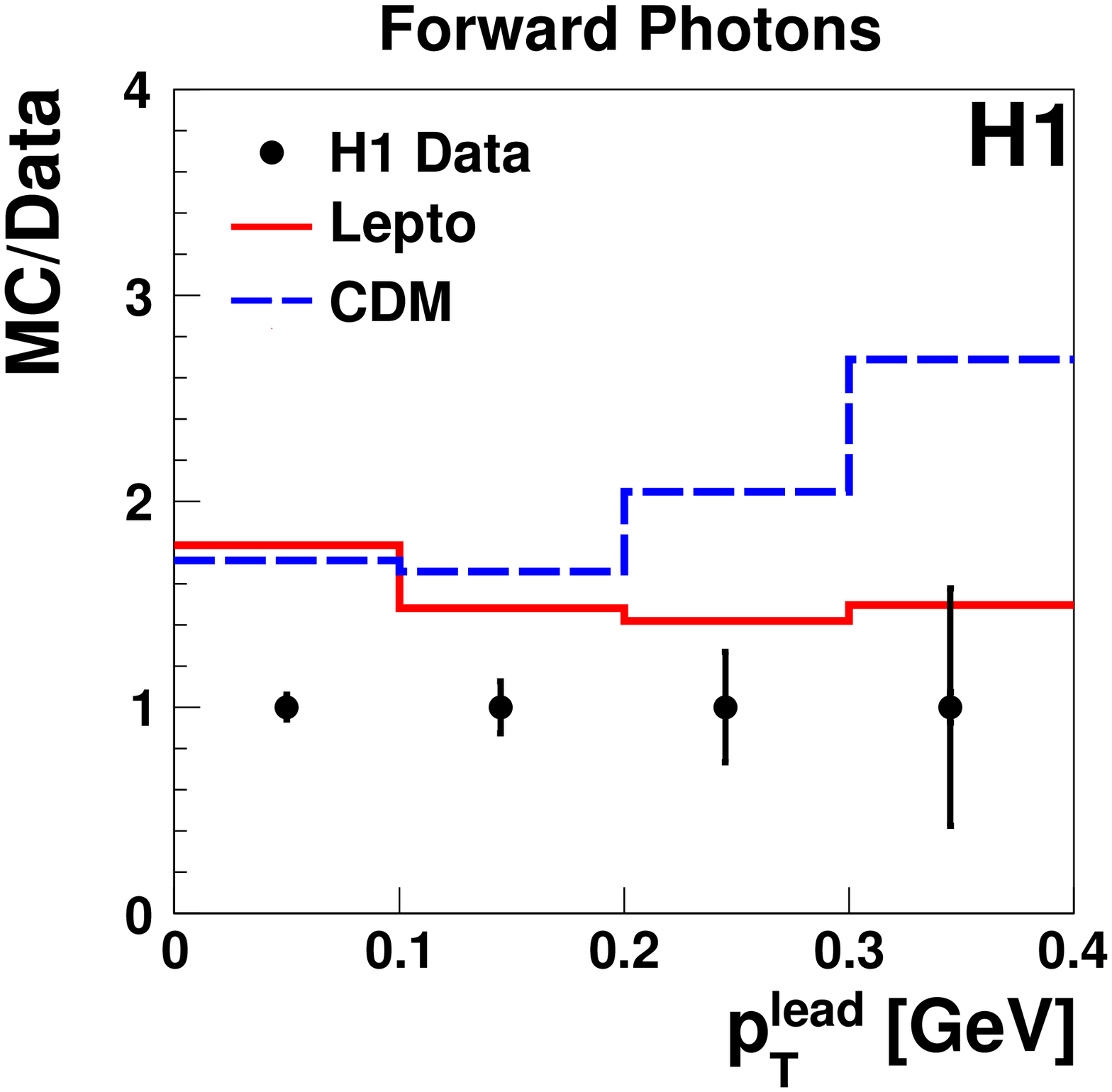,width=80mm}
\epsfig{file=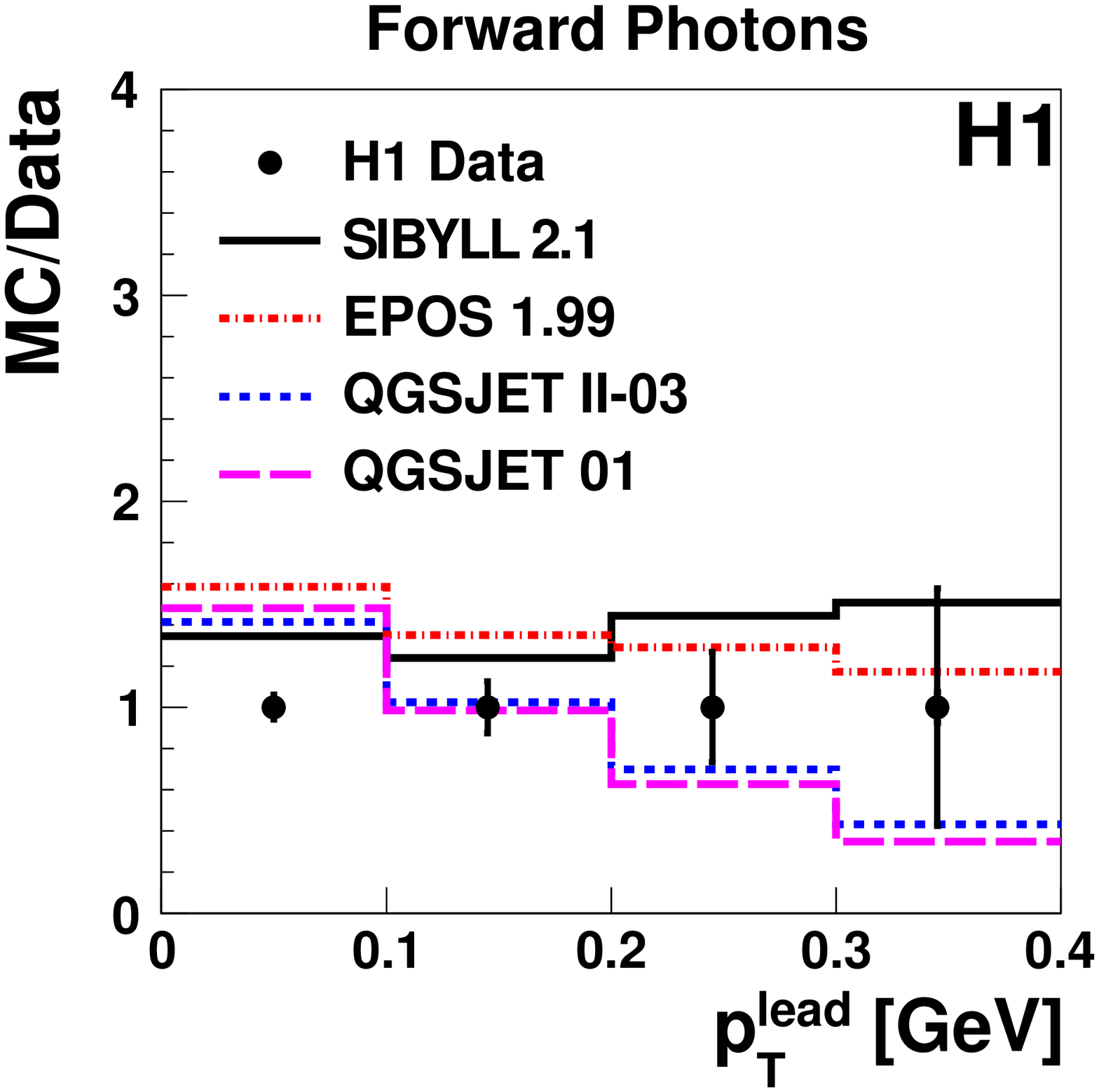,width=80mm}
\caption{
The normalised cross sections for the production of forward photons
in the pseudorapidity range $\eta>7.9$
in deep-inelastic scattering in the kinematic region $6< Q^2 < 100~$GeV$^2$ and $0.05<y<0.6$
as a function of the 
transverse momentum $p_T^{lead}$ of the leading photon
in the energy range $0.1<x_L^{lead}<0.7$.
The data are compared to two predictions of 
the DJANGOH Monte Carlo simulation, using LEPTO and CDM to simulate higher orders.
Also shown are models of
hadronic interactions, QGSJET, EPOS and SIBYLL.
The lower row shows the ratios of the Monte Carlo predictions
to the data. 
The error bars show the total experimental uncertainty, defined as the 
quadratic sum of the statistical and systematic uncertainties.
}
\label{fig:CrsecMCpt}
\end{figure}

\begin{figure}[p]
\epsfig{file=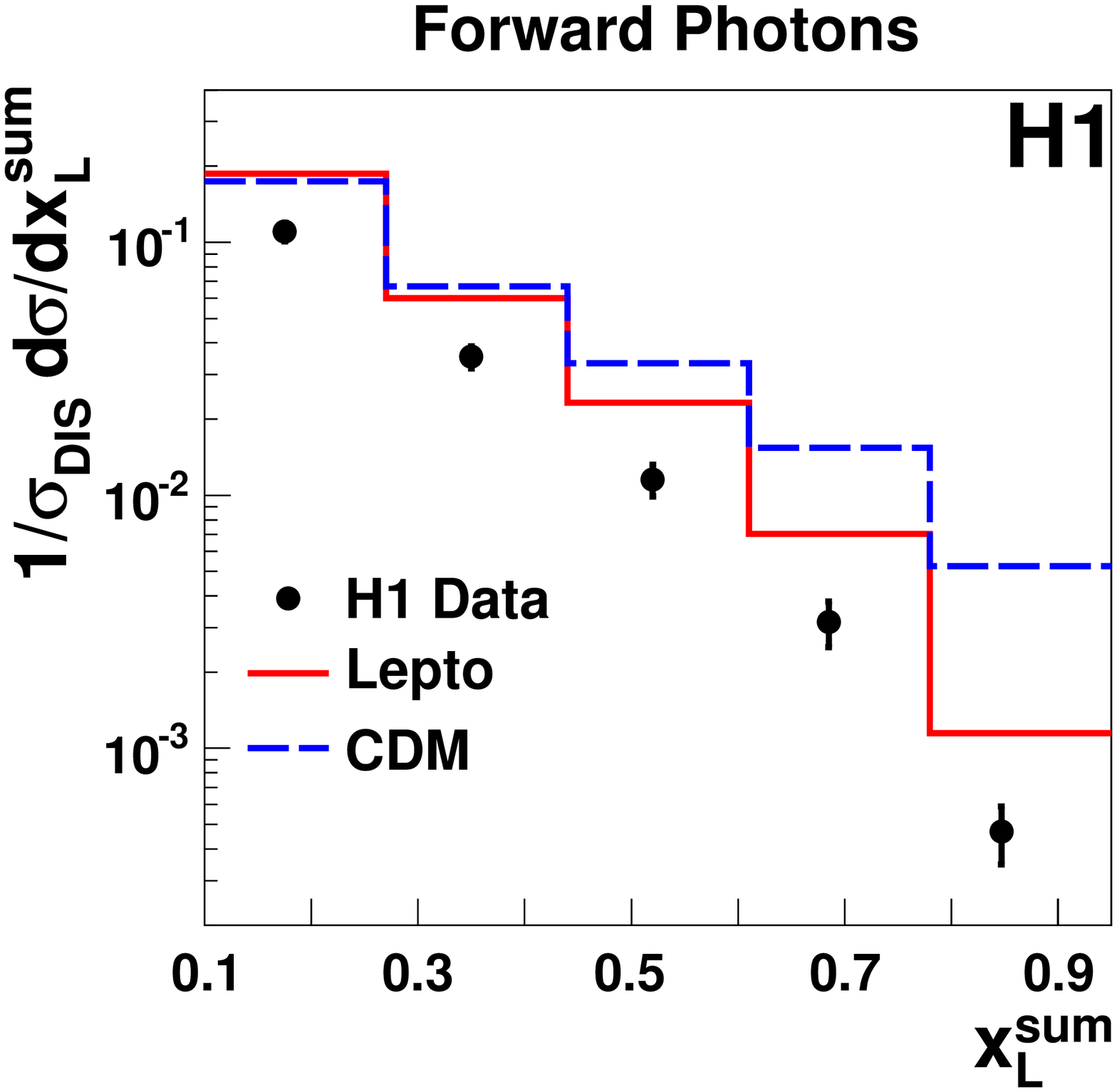,width=80mm}
\epsfig{file=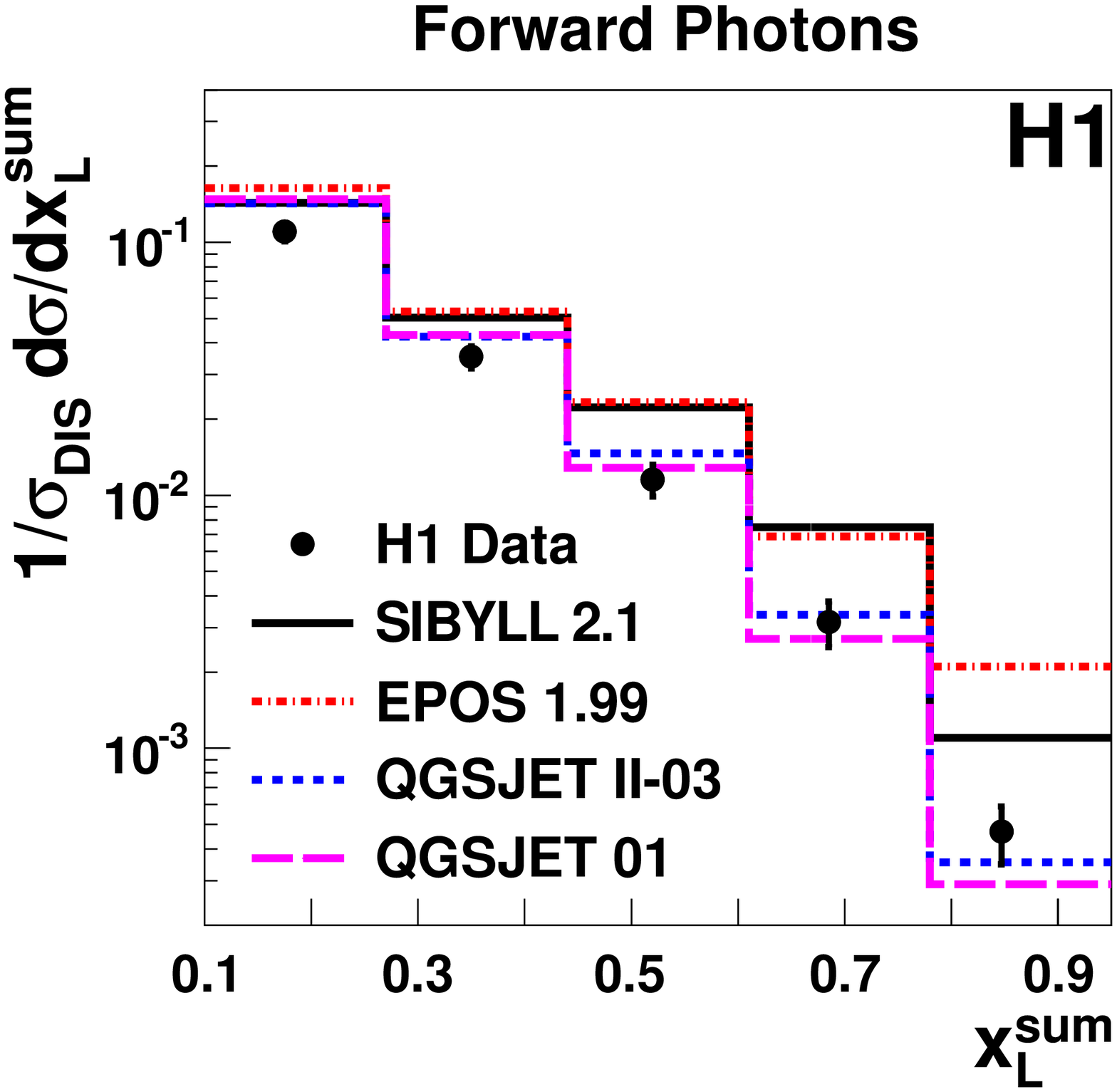,width=80mm}
\epsfig{file=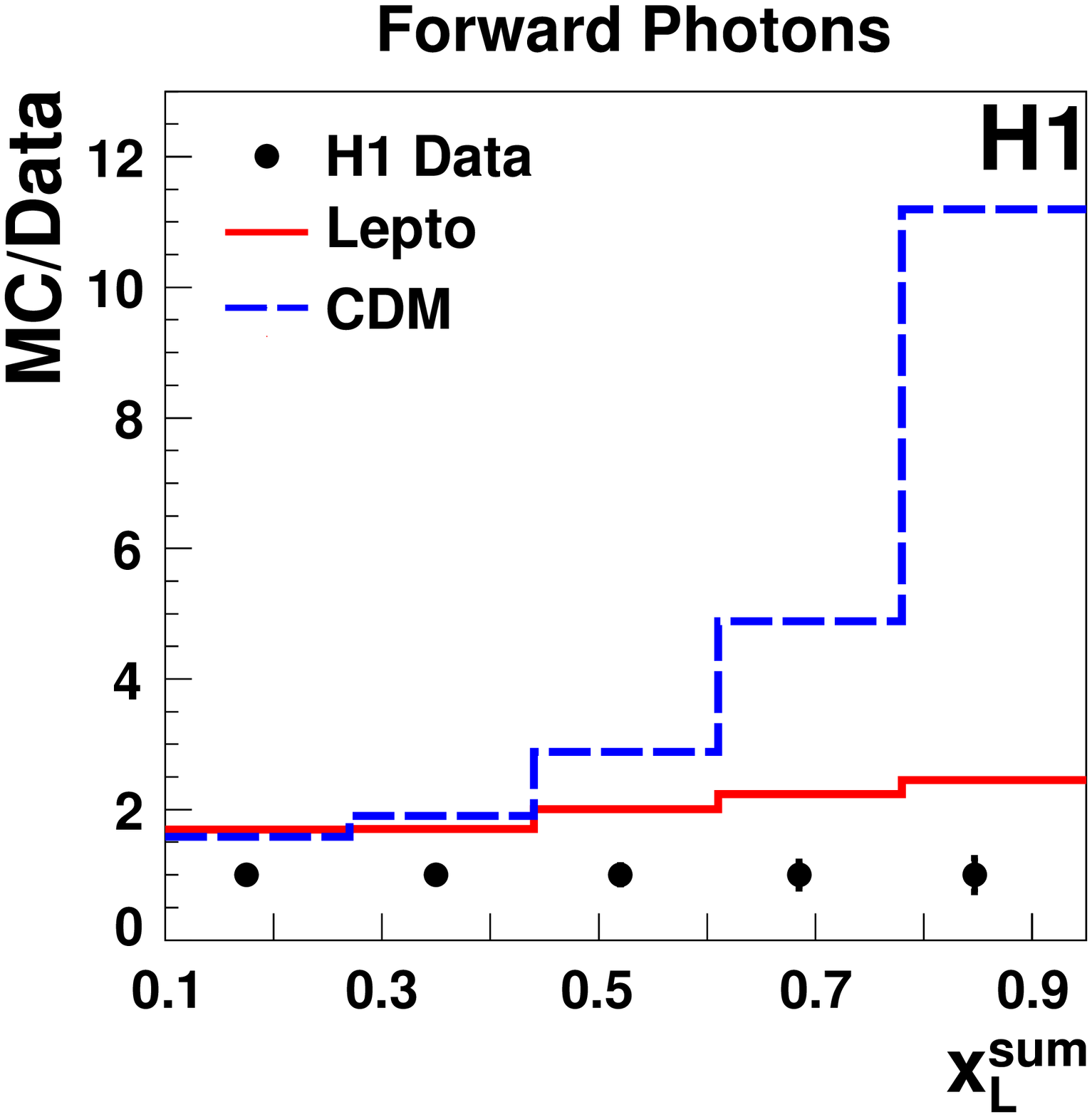,width=80mm}
\epsfig{file=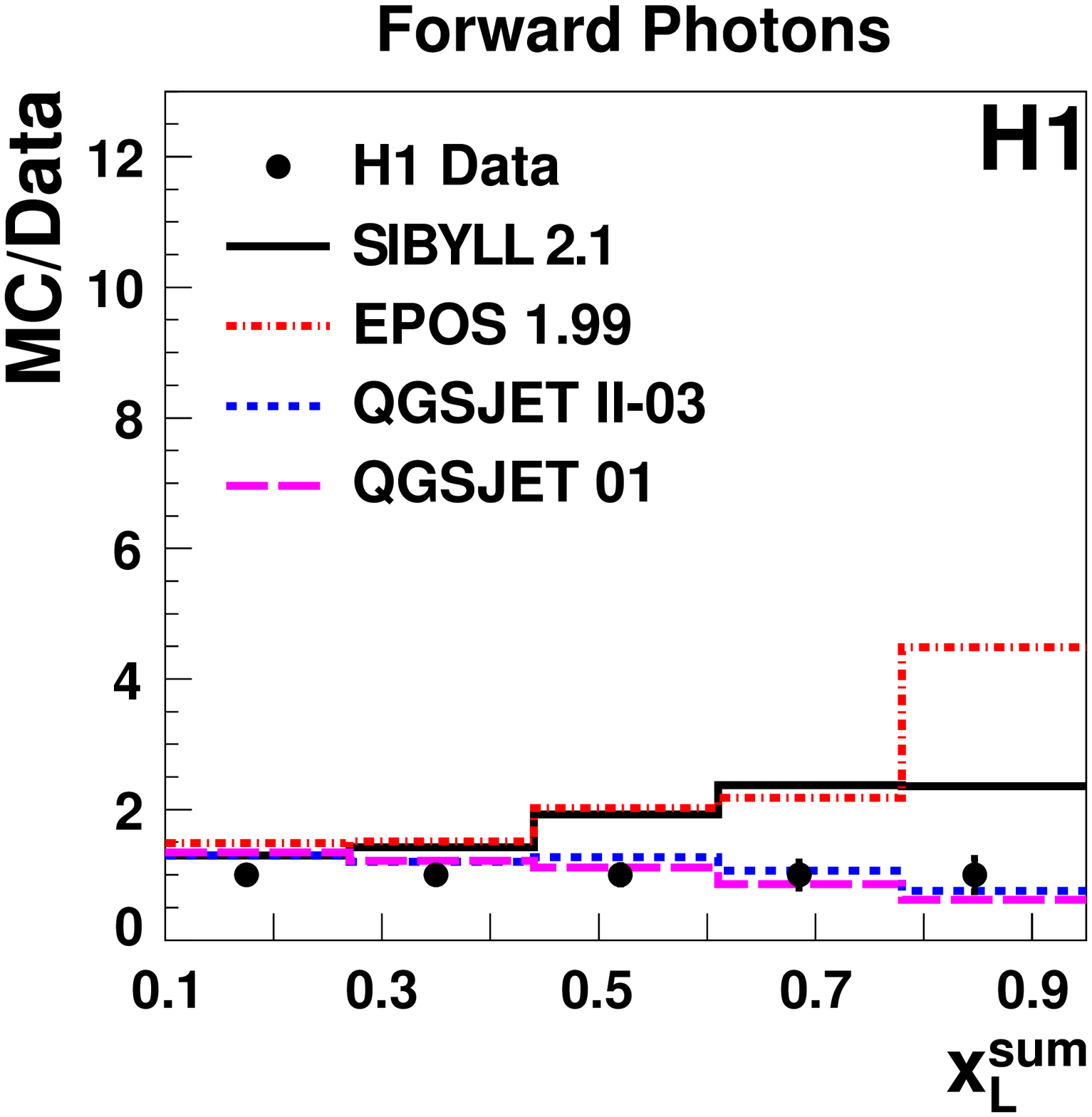,width=80mm}
\caption{
The normalised cross sections for the production of forward photons in the 
pseudorapidity range $\eta>7.9$
in deep-inelastic scattering in the kinematic region $6< Q^2 < 100~$GeV$^2$ and $0.05<y<0.6$
as a function of the sum of the longitudinal momentum fractions of photons $x_L^{sum}$.
The data are compared to two predictions of 
the DJANGOH Monte Carlo simulation, using LEPTO and CDM to simulate higher orders.
Also shown are models of
hadronic interactions, QGSJET, EPOS and SIBYLL.
The lower row shows the ratios of the Monte Carlo predictions to the data. 
The error bars show the total experimental uncertainty, defined as the 
quadratic sum of the statistical and systematic uncertainties.
}
\label{fig:CrsecMCsum}
\end{figure}

The measured normalised differential cross sections for the production
of very forward photons 
in the pseudorapidity range $\eta>7.9$
in DIS in the kinematic range $6<Q^2<100$~GeV$^2$ and 
$0.05<y<0.6$, are presented in table~\ref{tab:table1} and
figures~\ref{fig:CrsecMClead}-\ref{fig:CrsecMCsum}.
The measurements are presented in
figures~\ref{fig:CrsecMClead} and \ref{fig:CrsecMCpt} as a function of
$x_L^{lead}$ and $p_T^{lead}$ of the most energetic photon with
$0.1<x_L^{lead}<0.7$. 
The results as a function of the sum of
longitudinal momentum fractions $x_L^{sum}$ of all photons with
$\eta>7.9$ are presented in figure~\ref{fig:CrsecMCsum}.

The data are compared with the predictions of models for inclusive DIS
(LEPTO and CDM)
and models of hadronic interactions (EPOS, SIBYLL and
two versions of QGSJET).
The ratios of MC model predictions to the measurements are shown 
separately.

All models tested in this paper overestimate the total rate of forward
photons.  The LEPTO and CDM models predict about $70\%$ more photons
than measured, while EPOS, SIBYLL and QGSJET overestimate the rate of
photons by about $30\%$ to $50\%$.
In contrast to the excess of photons in the CDM, the same model 
predicts a too low rate of forward neutrons
as observed in previous H1 analysis of forward neutron production 
\cite{Aaron:2010ze}.

The shapes of all measured distributions are well described by
LEPTO. The CDM predicts harder $x_L$ and $p_T$ spectra.
The QGSJET model predicts slightly softer spectra in
$x_L$ and $p_T$. The EPOS and SIBYLL models predict harder
$x_L$ spectra, but describe reasonably the shape of $p_T$
distribution.

A measurement of  the energy spectra of single photons produced
at  $\eta>8.8$ in $pp$ collisions at $7$~TeV centre-of-mass energy  
at the LHC
has been recently reported by the LHCf Collaboration~\cite{Adriani:2011nf}.
This measurement also shows significant discrepancies with the
predictions from hadronic interaction models.
However, a direct comparison of the H1 and LHCf results is not possible
due to the different kinematic ranges of the two measurements.

\begin{figure}[pt]
\epsfig{file=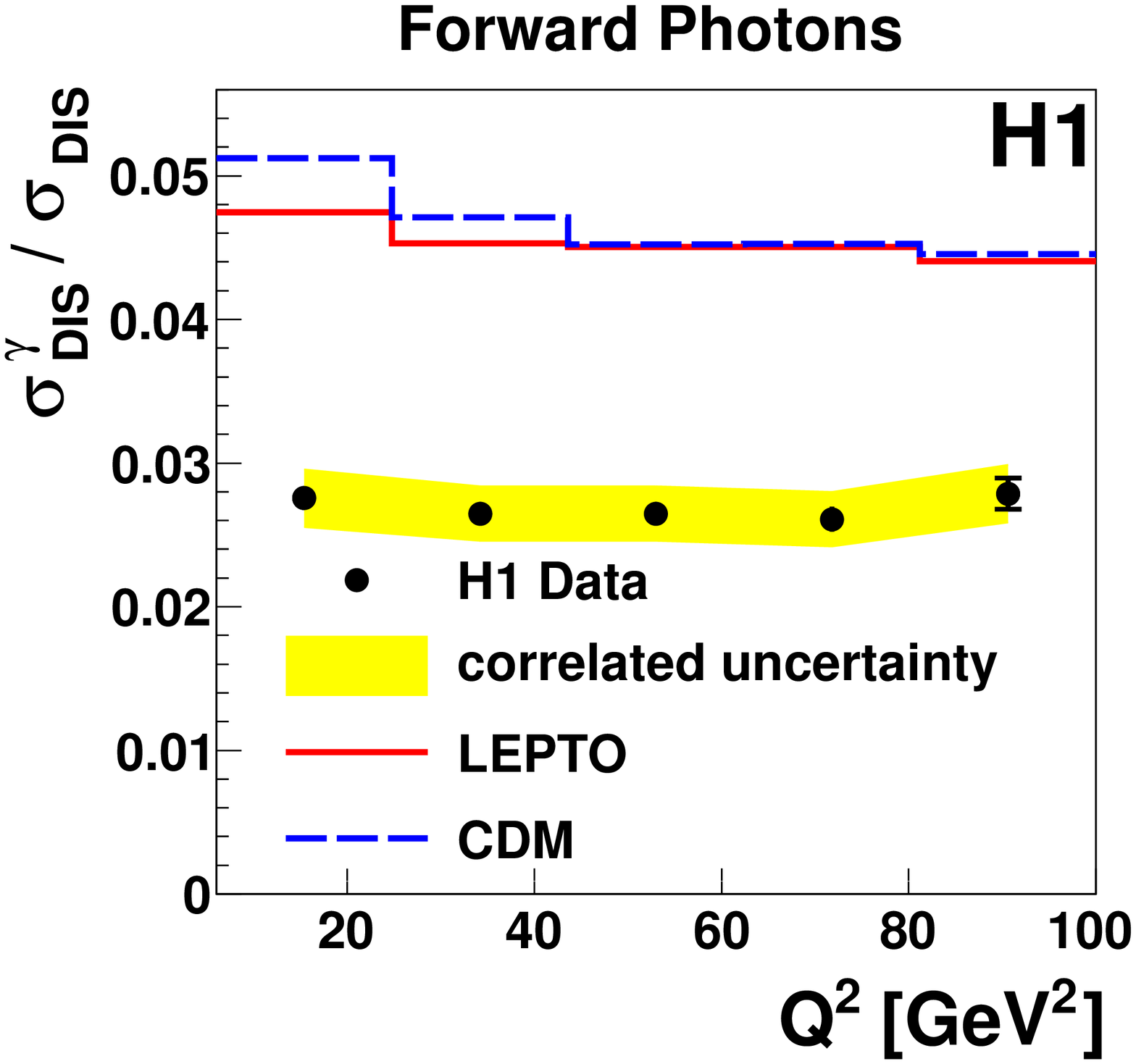,width=80mm}
\hspace*{3mm}
\epsfig{file=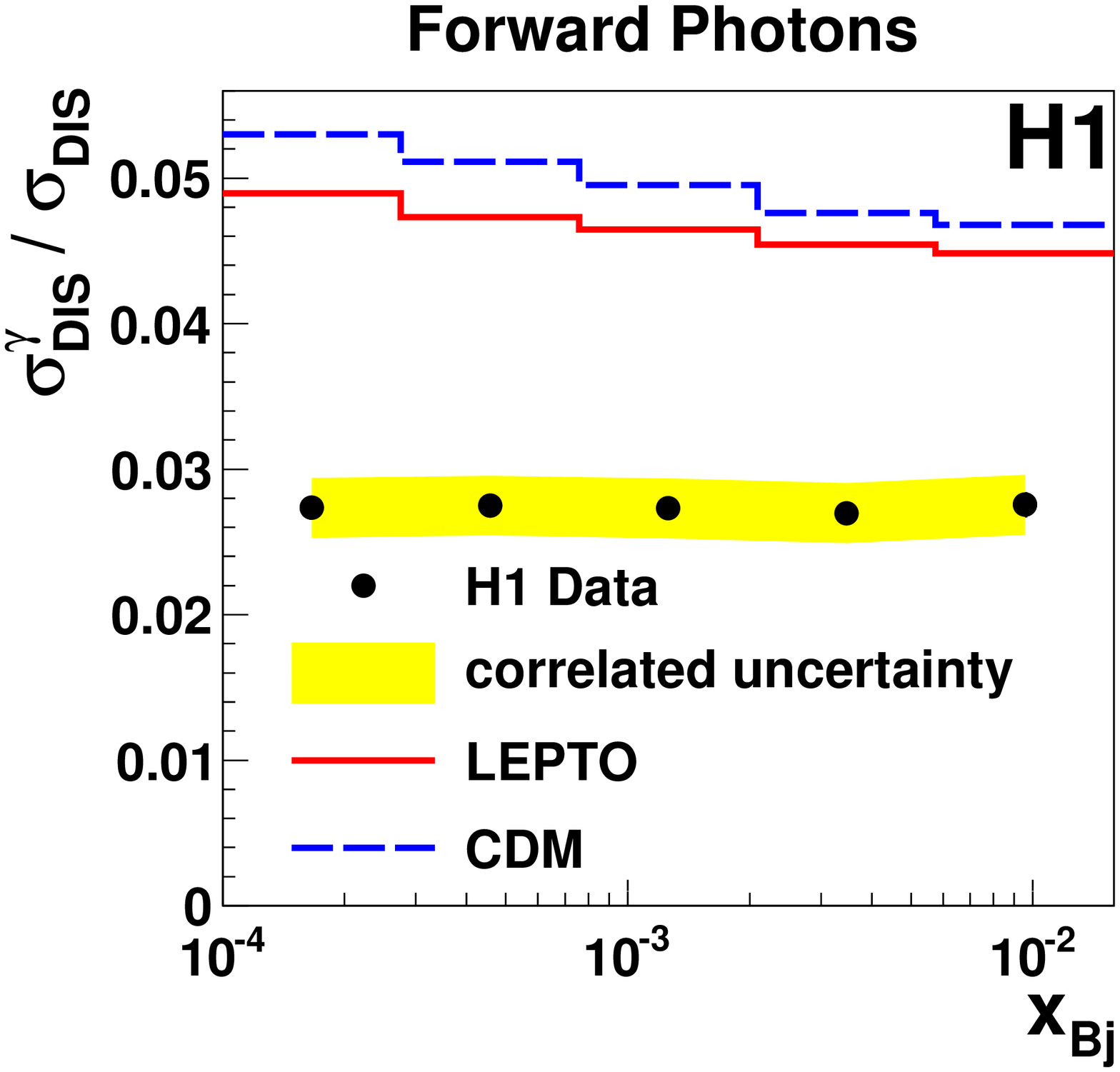,width=80mm}
\caption{The fraction of DIS events with forward photons
in the kinematic region $6< Q^2 < 100~$GeV$^2$ and $0.05<y<0.6$
and the pseudorapidity of the photon $\eta>7.9$
as a function of $Q^2$ and $x_{Bj}$.
The inner error bars shows the quadratic sum of the
statistical and the uncorrelated systematic uncertainties. 
Shaded band shows the correlated systematic uncertainties.
The expectation from the LEPTO and CDM models are also shown.
}
\label{fig:Q2xratio}
\end{figure}

The measurement of forward photons allows a test of the 
 limiting fragmentation hypothesis,  according to which 
the production of forward photons in DIS is insensitive to 
$Q^2$ and $x_{Bj}$.
To investigate this prediction, the ratio of the forward photon
production cross section to the inclusive DIS cross section
is measured as a function of $Q^2$ and $x_{Bj}$
(table~\ref{tab:table2} and figure~\ref{fig:Q2xratio}).
Within the uncertainties the fraction of DIS events with 
forward photons is independent from  $Q^2$ and $x_{Bj}$
in agreement with the limiting fragmentation hypothesis.
A similar conclusion was obtained in the earlier H1 analysis of forward
neutron production~\cite{Aaron:2010ze}.
The LEPTO and CDM predictions also included in 
%In 
figure~\ref{fig:Q2xratio} 
display a significant difference in normalisation compared to data as well as 
%also the LEPTO and CDM predictions 
% are included. Apart from the normalisation difference, 
a slight dependence as a function of $Q^2$ and $x_{Bj}$.

%--------------------------------------------------------------------------------------

\section{Summary}
\noindent
The production of high energy forward photons in the
pseudorapidity range $\eta>7.9$ is studied
for the first time at HERA
in deep-inelastic positron-proton scattering in the kinematic region
$6< Q^2 < 100$~GeV$^2$, $0.05<y<0.6$.
The normalised DIS cross sections are presented for the production of the most
energetic photon as a function of the longitudinal momentum fraction
and transverse momentum  in the range 
$0.1<x_L^{lead}<0.7$, and as a function  of the sum  of
longitudinal momentum fractions of
all forward photons in the range $0.1<x_L^{sum}<0.95$.
Predictions of Monte Carlo models overestimate the rate of photons.
The shapes of the measured cross sections are well described by the LEPTO
MC simulation, while the colour dipole model
predicts harder spectra in $x_L$ and $p_T$.
The measurement is also compared
to predictions of models  which are commonly used for the simulation of cosmic
ray air shower cascades.
All these models predict different spectra in $x_L$ and $p_T$.
None of the models can describe data in rate and in shape.
Within the measured kinematic range, the relative rate of forward photons 
in DIS events is observed to be independent of $Q^2$ and $x_{Bj}$,
in agreement with the hypothesis of limiting fragmentation.
The present measurement provides new information to further improve the understanding of proton
fragmentation in collider and cosmic ray experiments.

\section*{Acknowledgements}
We are grateful to the HERA machine group whose outstanding efforts 
have made this experiment possible. We thank the engineers and 
technicians for their work in constructing and maintaining 
the H1 detector, our funding agencies for financial support, 
the DESY technical staff for continual assistance and the DESY 
directorate for support and for the hospitality which they extend 
to the non-DESY members of the collaboration. We also wish to 
thank Tanguy Pierog and Ralph Engel for providing the predictions 
of cosmic ray models.

%--------------------------------------------------------------------------------------
%\bibliography{biblio}
%\bibliographystyle{f2lnart}

\newpage

\begin{table}[p]
\centering
{
\small
\begin{tabular}{|c|l|l|l||l|l|l|l|}
\hline  
& & & & & \multicolumn{3}{|c|}{\underline{\hspace*{0.1cm} correlated sys. uncertainty\hspace*{0.1cm}} } \\
$x_L^{lead}$ range  &  {\Large$\frac{1}{\sigma_{DIS}}\frac{d\sigma}{d x_L^{lead}}$} & 
$\delta_{stat.}$ & $\delta_{total~sys.}$ &  $\delta_{uncorrel.sys.}$  &  
$\delta_{E_{FNC}}$ & $\delta_{XY_{FNC}}$ & $\delta_{model}$   \\
& & & & &  & & \\ [-5pt]
\hline 
&  & & & & & & \\ [-12pt]
$ 0.10 \div 0.22 $  &  \hspace{5mm}  $0.134$  & $0.001$    & $0.011$    & $0.002$   & $0.001$   & $0.011$    & $0.001$   \\ 
$ 0.22 \div 0.34 $  &  \hspace{5mm}  $0.0577$ & $0.0005$   & $0.0061$   & $0.0012$  & $0.0029$  & $0.0052$   & $0.0008$  \\
$ 0.34  \div 0.46 $ &  \hspace{5mm}  $0.0226$ & $0.0003$   & $0.0029$   & $0.0005$  & $0.0018$  & $0.0023$   & $0.0003$  \\
$ 0.46  \div 0.58 $ &  \hspace{5mm}  $0.00764$ & $0.00017$ & $0.00123$  & $0.00029$ & $0.00061$ & $0.00099$  & $0.00027$   \\
$ 0.58  \div 0.70 $ &  \hspace{5mm}  $0.00229$ & $0.00008$ & $0.00048$  & $0.00017$ & $0.00025$ & $0.00034$  & $0.00016$  \\ 
\hline
& & & & & \multicolumn{3}{|c|}{\underline{\hspace*{0.1cm} correlated sys. uncertainty\hspace*{0.1cm}} } \\
$p_T^{lead}$ range  &  {\Large$\frac{1}{\sigma_{DIS}}\frac{d\sigma}{d p_T^{lead}}$} & 
$\delta_{stat.}$ & $\delta_{total~sys.}$ &  $\delta_{uncorrel.sys.}$  & 
$\delta_{E_{FNC}}$ & $\delta_{XY_{FNC}}$ & $\delta_{model}$   \\
& & & & & & &  \\ [-12pt]
$[$GeV$]$ &  \hspace*{5mm}   $[$GeV$^{-1}]$ &    $[$GeV$^{-1}]$ &    $[$GeV$^{-1}]$ &   
 $[$GeV$^{-1}]$ & $[$GeV$^{-1}]$ &    $[$GeV$^{-1}]$ & $[$GeV$^{-1}]$ \\
& & & & & & &  \\ [-8pt]
\hline
& & &  & & & & \\ [-12pt]
$ 0.0 \div 0.1 $ &  \hspace{5mm} $0.159$   & $0.001$   & $0.010$   &  $0.003$   & $0.005$   & $0.008$   & $0.001$  \\
$ 0.1 \div 0.2 $ &  \hspace{5mm} $0.0971$  & $0.0010$  & $0.0116$  &  $0.0041$  & $0.0068$  & $0.0078$  & $0.0034$   \\
$ 0.2 \div 0.3 $ &  \hspace{5mm} $0.0220$  & $0.0005$  & $0.0056$  &  $0.0010$  & $0.0024$  & $0.0048$  & $0.0008$  \\
$ 0.3 \div 0.4 $ &  \hspace{5mm} $0.00395$ & $0.00029$ & $0.00229$ &  $0.00025$ & $0.00087$ & $0.00209$ & $0.00020$  \\ 
\hline
& & & & & \multicolumn{3}{|c|}{\underline{\hspace*{0.1cm} correlated sys. uncertainty\hspace*{0.1cm}} } \\
$x_L^{sum}$ range  &  {\Large$\frac{1}{\sigma_{DIS}}\frac{d\sigma}{d x_L^{sum}}$} & 
$\delta_{stat.}$ & $\delta_{total~sys.}$ &  $\delta_{uncorrel.sys.}$  & 
$\delta_{E_{FNC}}$ & $\delta_{XY_{FNC}}$ & $\delta_{model}$  \\
& & & & & & &   \\ [-8pt]
\hline 
&  & & & & & & \\ [-12pt]
$ 0.10 \div 0.27 $  & \hspace{5mm} $0.110$    & $0.001$    & $0.009$    & $0.002$    & $0.001$    & $0.009$    & $0.002$  \\ 
$ 0.27 \div 0.44 $  & \hspace{5mm} $0.0353$   & $0.0003$   & $0.0038$   & $0.0009$   & $0.0018$   & $0.0032$   & $0.0007$  \\
$ 0.44 \div 0.61 $  & \hspace{5mm} $0.0115$   & $0.00021$  & $0.0018$   & $0.0007$   & $0.0009$   & $0.0012$   & $0.0006$  \\ 
$ 0.61 \div 0.78 $  & \hspace{5mm} $0.00315$  & $0.00011$  & $0.00068$  & $0.00032$  & $0.00032$  & $0.00041$  & $0.00031$  \\
$ 0.78 \div 0.95 $  & \hspace{5mm} $0.000468$ & $0.000039$ & $0.000172$ & $0.000050$ & $0.000140$ & $0.000070$ & $0.000050$  \\
\hline
\end{tabular}
\normalsize

}
\caption{
The normalised cross sections for the production of forward photons
in the pseudorapidity range $\eta>7.9$
in deep-inelastic scattering in the kinematic
region $6< Q^2 < 100$~GeV$^2$ and $0.05<y<0.6$
as a function of the longitudinal momentum fraction $x_L^{lead}$ and
transverse momentum $p_T^{lead}$ of the most energetic 
photon in the energy range $0.1<x_L^{lead}<0.7$ 
and as a function of the sum of the longitudinal momentum fractions
of photons $x_L^{sum}$.
For each measurement, the  statistical, the total systematic, the
uncorrelated systematic uncertainties,  and the bin-to-bin correlated
systematic uncertainties due to the FNC absolute energy scale, 
the impact position of the FNC and the model dependence of data 
correction are given.
}
\label{tab:table1}
\end{table}

\newpage

\begin{table}[p]
\centering
{
\small
\begin{tabular}{|c|c|c|c||c|c|c|c|}
\hline 
& & & & & \multicolumn{3}{|c|}{\underline{\hspace*{0.1cm} correlated sys. uncertainty\hspace*{0.1cm}} } \\
$Q^2$ range $[GeV^2]$  &  { \Large$\frac{\sigma^\gamma_{DIS}(Q^2)}{\sigma_{DIS}(Q^2)}$} & $\delta_{stat.}$ & 
$\delta_{total~sys.}$ &  $\delta_{uncorrel.sys.}$  &  
$\delta_{E_{FNC}}$ & $\delta_{XY_{FNC}}$ & $\delta_{model}$   \\
&  & & & & & & \\ [-5pt]
\hline 
&  & & & & & & \\ [-12pt]
$ ~~6.0  \div ~~24.8 $  &  $0.0276 $ & $0.0001$ & $0.0020$ & $0.0003$ & $0.0011$ & $0.0017$ & $0.0001$  \\ 
$  24.8  \div ~~43.6 $  &  $0.0265 $ & $0.0003$ & $0.0020$ & $0.0003$ & $0.0011$ & $0.0016$ & $0.0001$ \\
$  43.6  \div ~~62.4 $  &  $0.0265 $ & $0.0005$ & $0.0020$ & $0.0004$ & $0.0011$ & $0.0016$ & $0.0001$ \\
$  62.4  \div ~~81.2 $  &  $0.0261 $ & $0.0007$ & $0.0020$ & $0.0005$ & $0.0010$ & $0.0016$ & $0.0001$ \\
$  81.2  \div 100.0 $   &  $0.0279 $ & $0.0011$ & $0.0021$ & $0.0005$ & $0.0011$ & $0.0017$ & $0.0001$ \\ 
\hline
& & & & & \multicolumn{3}{|c|}{\underline{\hspace*{0.1cm} correlated sys. uncertainty\hspace*{0.1cm}} } \\
$x_{Bj}$ range &  { \Large$\frac{\sigma^\gamma_{DIS}(x_{Bj})}{\sigma_{DIS}(x_{Bj})}$}  & $\delta_{stat.}$ & 
$\delta_{total~sys.}$ &  $\delta_{uncorrel.sys.}$  &  
$\delta_{E}$ & $\delta_{XY}$ & $\delta_{model}$   \\
&  & & & & & & \\ [-5pt]
\hline
&  & & & & & & \\ [-12pt]
$ 1.00 \cdot 10^{-4} \div 2.75 \cdot 10^{-4} $ &  $0.0273$ & $0.0003$ & $0.0020$ & $0.0004$ & $0.0011$ & $0.0016$ & $0.0001$  \\
$ 2.75 \cdot 10^{-4} \div 7.69 \cdot 10^{-4} $ &  $0.0275$ & $0.0002$ & $0.0020$ & $0.0003$ & $0.0011$ & $0.0017$ & $0.0001$  \\
$ 7.69 \cdot 10^{-4} \div 2.98 \cdot 10^{-3} $ &  $0.0273$ & $0.0002$ & $0.0020$ & $0.0004$ & $0.0011$ & $0.0016$ & $0.0001$  \\
$ 2.98 \cdot 10^{-3} \div 5.75 \cdot 10^{-3} $ &  $0.0270$ & $0.0003$ & $0.0020$ & $0.0004$ & $0.0011$ & $0.0016$ & $0.0001$   \\
$ 5.75 \cdot 10^{-3} \div 1.58 \cdot 10^{-2} $ &  $0.0276$ & $0.0007$ & $0.0021$ & $0.0006$ & $0.0011$ & $0.0017$ & $0.0001$  \\ 
\hline
\end{tabular}
\normalsize

}
\caption{
The fraction of DIS events with forward photons
in the kinematic region $6< Q^2 < 100$~GeV$^2$ and $0.05<y<0.6$
and the pseudorapidity of the photon $\eta>7.9$.
For each measurement, the  statistical, the total systematic, the
uncorrelated systematic uncertainties,  and the bin-to-bin correlated
systematic uncertainties due to the FNC absolute energy scale, 
the impact position of the FNC and the model dependence of data 
correction are given.
}
\label{tab:table2}
\end{table}

%%%%%%%%%%%%%%%%%%%%%%%%%%%%%%%%%%%%%%%%%%%%%%%%%%%%%%%%%%%%%%%
\newpage

\end{document}